\documentclass{aa} 

\usepackage{graphicx} 
\usepackage{txfonts} 
\usepackage{natbib} 
 
\newcommand{\Mpc}{h^{-1}\, {\rm Mpc}}

\newcommand{\be}{\begin{equation}}
\newcommand{\ee}{\end{equation}}
\newcommand{\bea}{\begin{equation}\begin{aligned}} 
\newcommand{\eea}{\end{aligned}\end{equation}}

\newcommand{\dd}[1]{{\rm d}#1\,}

\def\apj{ApJ} 
\def\apjl{ApJL} 
\def\apjs{ApJS} 
 
\def\aap{A\&A} 
\def\mnras{MNRAS}

\usepackage{amstext}

\begin{document}    
 
\title{Evolution of skewness and kurtosis of cosmic density fields}

\author{Jaan Einasto\inst{1,2,3} 
\and Anatoly Klypin\inst{4,5}
 \and  Gert H\"utsi\inst{6}
 \and  Lauri-Juhan Liivam\"agi\inst{1}
\and Maret Einasto\inst{1}   
}
\institute{Tartu Observatory, University of Tartu, EE-61602 T\~oravere, Estonia 
\and  
ICRANet, Piazza della Repubblica 10, 65122 Pescara, Italy 
\and 
Estonian Academy of Sciences, 10130 Tallinn, Estonia
\and
Astronomy Department, New Mexico State University, Las Cruces, NM 88003, USA
\and
Department of Astronomy, University of Virginia, Charlottesville, VA 22904, USA
\and
National Institute of Chemical Physics and Biophysics, 
Tallinn 10143, Estonia  
}

\date{ Received 26 November 2020; accepted 25 June 2021}  
 
\authorrunning{Einasto et al.} 
 
\titlerunning{Evolution of skewness and kurtosis} 
 
\offprints{Jaan Einasto, e-mail: jaan.einasto@ut.ee} 
 
\abstract {}
{We investigate the evolution of the one-point probability distribution
  function (PDF) of the dark matter density field and the
  evolution of its moments for fluctuations that are  Gaussian in the
  linear regime.}
{We performed numerical simulations of the evolution of the cosmic web
  for the conventional $\Lambda$CDM model. The simulations covered a
  wide range of box sizes $L=256-4000~\Mpc$, mass, and force resolutions,
  and epochs from very early moments $z=30$ to the present moment
  $z=0$. We calculated density fields with various smoothing lengths to
  determine the dependence of the density field on the smoothing scale.  We
  calculated the PDF and its moments variance, skewness, and
  kurtosis. We determined the dependence of  these parameters on the
  evolutionary epoch $z$, on the smoothing length $R_t$, and on the
  rms deviation of the density field $\sigma$ using a cubic-cell and
  top-hat smoothing with kernels $0.4~\Mpc \le R_t \le 32~\Mpc$.}
{We focus on the third (skewness $S$) and fourth (kurtosis $K$)
  moments of the distribution functions: their dependence on the
  smoothing scale $R_t$, the amplitude of the fluctuations $\sigma$ , and
  the redshift $z$.  Moments $S$ and $K$, calculated for density
  fields at different cosmic epochs and smoothed with various scales,
  characterise the evolution of different structures of the web.
  Moments calculated with small-scale smoothing
  ($R_t\approx (1-4)~\Mpc$) characterise the evolution of the web on
  cluster-type scales.  Moments found with strong smoothing
  ($R_t\gtrsim (5-15)~\Mpc$) describe the evolution of the web on
  supercluster scales.  During the evolution, the reduced skewness
  $S_3= S/\sigma$ and reduced kurtosis $S_4=K/\sigma^2$ present a
  complex behaviour: at a fixed redshift, curves of $S_3(\sigma)$ and
  $S_4(\sigma)$ steeply increase with $\sigma$ at $\sigma\lesssim 1$
  and then flatten out and become constant at $\sigma\gtrsim 2$.  When
  we fixed the smoothing scale $R_t$, the curves at large $\sigma$ started to gradually
  decline after reaching the maximum at
  $\sigma\approx 2$, . We provide accurate fits for the evolution of
  $S_{3,4}(\sigma,z)$. Skewness and kurtosis approach constant levels
  at early epochs $S_3(\sigma) \approx 3$ and
  $S_4(\sigma) \approx 15$. }
  {Most of the statistics of dark matter clustering (e.g. halo mass
    function or concentration-mass relation) are nearly universal:
    they mostly depend on the $\sigma$ with a relatively modest
    correction to apparent dependence on the redshift. We find just
    the opposite for skewness and kurtosis: the dependence of the moments on
  the evolutionary epoch $z$ and smoothing length $R_t$ is very different.
  Together, they uniquely determine the evolution of $S_{3,4}(\sigma)$ .
     The evolution of $S_3$ and $S_4$ cannot
    be described by current theoretical approximations. The often used
    lognormal distribution function for the PDF fails to even
    qualitatively explain the shape and evolution of $S_3$ and $S_4$.}

\maketitle

\section{Introduction} 

According to the currently accepted cosmological paradigm, the
evolution of the structure in the Universe began from small
perturbations that were created during the epoch of inflation.  The
structure evolved by gravitational amplification to form the cosmic
web that is observed now.  It is also accepted that initial density
fluctuations were random (but correlated) and had a Gaussian
distribution.  { The Gaussian random field is symmetrical around the
mean density, that is,  positive and negative deviations from the mean
density are equally probable.}  On the other hand, it is well known that the
current density field of the cosmic web is highly asymmetric: positive
density departures from the mean density can be very strong, while the
negative deviations are restricted by the condition that the density
cannot be negative.  The asymmetry of the density field can be studied
with a one-point probability distribution function (PDF) of the density
field and its moments.

There are different ways of studying the properties of PDFs.  Analytical methods are one approach
\citep{Peebles:1980aa,Bernardeau:1995aa,Bernardeau:2002aa}. In this
case, the PDF is modelled theoretically using cosmological perturbation
theory (PT), which allows calculating the PDF and its moments variance,
skewness, and kurtosis. The basic elements of the cosmological PT and its
applications were discussed in detail by \citet{Peebles:1980aa},
\citet{Bernardeau:2002aa}, and \citet{Szapudi:2009aa}.  Another
possibility is calculating the evolution of the PDF numerically
using N-body simulations. For early studies, see \citet{Kofman:1992aa} and
\citet{Kofman:1994aa}.

The asymmetry and flatness of the PDF are measured by the third
(skewness $S$) and fourth (kurtosis $K$) moments of the distribution
functions. The moments are the most simple forms of the three-point and
four-point correlation functions, and they therefore cannot be reduced
to second-order statistics such as the correlation function or the 
power spectrum.  In mathematical statistics, skewness and kurtosis of
 a random variable are 
defined as dimensionless parameters and can be called mathematical
skewness $S$ and mathematical kurtosis $K$. They change during the
evolution and can be used to characterise the evolution.  In cosmology, there is a
tradition to define skewness and kurtosis in a different way.  These
skewness and kurtosis parameters are called reduced 
\citep{Lahav:1993aa}.   { To emphasise the difference between
  mathematical and cosmological terminology, we mostly use the term ``cosmological''. 
  Early studies suggested that during the evolution, cosmological
  skewness $S_3$ and cosmological kurtosis $S_4$ remained approximately
  constant and that they characterise the general properties of the model of the
  universe \citep{Peebles:1980aa}}.

Simple relations exist between mathematical and cosmological
parameters. The skewness is $S(\sigma) = S_3 \times\sigma$, and the kurtosis is
$K(\sigma) = S_4 \times\sigma^2$, where $\sigma$ is the standard
deviation of fluctuations of the density field.  The initial density
field that is generated during the inflation must have a density fluctuation
with finite non-zero amplitude, $\sigma > 0$.  Moreover,
if initial fluctuations were Gaussian,
then they should be symmetrical.  Thus the question is how 
 the asymmetry in the density distribution forms and evolves.

The tradition of quantifying the moments of the PDF follows
\citet{Peebles:1980aa}.  Based on the linear PT, Peebles found that for
the Einstein-de Sitter model with $\Omega=1,$ the cosmological skewness
has the value $S_3=34/7$.  Later studies showed that $S_3$ also
depends on the effective index $n$ of the power spectrum, 
$P(k) \propto k^n$, as well as on the smoothing length $R$
\citep{Bouchet:1992aa, Bouchet:1992ab, Juszkiewicz:1993aa,
  Bernardeau:1994aa}.  Subsequent studies of the PDF and its moments have
confirmed and extended these results; see for example
\citet{Catelan:1994aa}, \citet{Bernardeau:1995aa},
\citet{Juszkiewicz:1995aa}, \citet{Lokas:1995aa},
\citet{Gaztanaga:1998aa}, \citet{Gaztanaga:2000aa},
\citet{Kayo:2001aa}, and \citet{Uhlemann:2017ad}.  These studies were
theoretical and used various methods of the perturbation theory to
follow the evolution of PDFs of the density field and its moments.

Various theoretical approximations were suggested to determine the values of
cosmological skewness and kurtosis parameters.
\citet{Bernardeau:1995aa} listed in their Table~1 the values of cosmological
skewness $S_3$ and kurtosis $S_4$ for various approximations.
Depending on the approximation, the $S_3$ and $S_4$ values depend
differently on the index $n$ of the power spectrum. Using analytical
methods, the authors calculated the dependence of the PDF moments on $\sigma$
and on the high-end cutoff of the PDF.

{ \citet{Kofman:1994aa} were one of the first to investigate the
  evolution of one-point distributions from Gaussian initial
  fluctuations using numerical simulations. The authors approximated the
  evolution by the Zeldovich formalism and found that the PDF of the
  density field rapidly obtains a log-normal shape. They also noted
  that the moments of the density distribution gradually deviate from
  Gaussian in the whole range of $\sigma$ they tested. On the basis of
  numerical simulations, the authors calculated PDFs for various epochs and
  smoothing radii; see Figure~5 in \citet{Kofman:1994aa}. The basic data
  of their model, as well as of other models that were based on numerical
  simulations, are given in Table~\ref{Tab3} below.}

\citet{Marinoni:2008aa} used the  Visible Multi-Object
Spectrograph Very Large Telescope    (VIMOS VLT)
Deep Survey by \citet{Marinoni:2005ut} over the redshift range $0.7 <
z < 1.5$ at a scale $R=10~\Mpc$ and reported that the skewness decreases
with increasing redshift  $2.0 \ge S_3 \ge 1.3$, in good agreement
with the prediction by \citet{Fry:1993aa}.  
\citet{Romeo:2008aa} investigated discreteness effects in 
cosmological constant $\Lambda$ cold dark matter  ($\Lambda$CDM)
simulations and their effects on cosmological parameters such as the standard
deviation $\sigma$, the skewness $S$, and the kurtosis $K$.  For the present
epoch, $z=0$, $\sigma \approx 20$, $S \approx 120$, and
$K \approx 2.5\times10^4$.  These high values are expected for a
smoothing kernel of size $R_t \ll 1~\Mpc$.

\citet{Hellwing:2009wx} used numerical simulations to investigate the
role of long-range scalar interactions in the DM model. As tests, the 
authors studied the power spectrum, the correlation function, and
the PDF of various models. 
\citet{Hellwing:2010aa} used a series of N-body simulations to test
the $\Lambda$CDM and a modified dark matter (DM) model. Models were compared
using cosmological moments of 
the density field, $S_3, \dots, S_8$.  \citet{Hellwing:2010wp} studied
the effect of long-range scalar DM interactions on properties of
galactic haloes.   \citet{Hellwing:2017aa} continued the
comparison of the $\Lambda$CDM and the modified gravity models with mild and
strong growth-rate enhancement.  

\citet{Pandey:2013aa} investigated the evolution of the density field
of the Millennium and Millennium II simulations. The density distribution in
the Millennium simulations is shown in Figure 4 of \citet{Pandey:2013aa}. 
\citet{Mao:2014aa} used $N$-body simulations to investigate whether
measurements of the PDF moments can yield constraints on primordial
non-Gaussianity.  The authors reported a dependence of the standard deviation
$\sigma$, cosmological skewness $S_3$ , and cosmological kurtosis $S_4$
using smoothing radii $R_t=10 - 100~\Mpc$. All moments decrease with
increasing smoothing length $R_t$. For smoothing spheres of radii 
$R_t=10 - 100~\Mpc,$\ the authors found $2.5 \le S_3 \le 4$ and
$5 \le S_4 \le 30$.

\citet{Shin:2017aa} found a new fitting formula for the PDF that
describes the density distribution better than the log-normal
formula. The parameters of the fitting formula were determined on the basis of
numerical simulations for various input cosmological simulations in the
interval of cosmic epochs $z \le 4$. 

{ In spite of the  extensive literature on the subject, there are
  important aspects that have not been sufficiently studied so
  far. Most of the attention was 
paid to the shape of the PDF at redshift $z=0$. The evolution of the
PDF with redshift were studied, and it was investigated whether the
evolution of PDF is related  only to  changes 
in the amplitude of fluctuations $\sigma(z)$. This can be true in
the linear regime, but it remains to be determined what occurs in the non-linear stage. }

The goal of this study is to investigate the evolution of the
cosmological density distribution function and its moments and to
determine the relations between the parameters defined by mathematical and
cosmological methods.  { We use $N$-body simulations to study the
  evolution of the PDF}.  We assume that seeds of the cosmic web were 
created by initially small fluctuations of the early universe in the inflationary
phase, and that these fluctuations had a Gaussian distribution.  
We also assume that the currently accepted  
  $\Lambda$CDM model represents
the actual universe accurately enough and that it can be used to investigate the
evolution of the structure of the real universe.

The paper is organised as follows. In Section~2 we describe
the numerical simulations we used and the methods with which we calculated the density field,
the PDF of density fields, and the method we used to determine its moments, the
variance, skewness, and kurtosis. {  In Section 3 we describe and analyse basic
results for various  redshifts and smoothing lengths.}  In
 Section 4 we discuss our results and compare numerical
results with the evolving pattern of the density field.   The  last
section  contains our conclusions.

\section{Data and methods}

In this section we describe our simulations of the evolution of the
cosmic web and calculate the density field, its PDF, and its moments. Our
emphasis is on describing the connections between the statistical  and
cosmological definitions of skewness and kurtosis.

We calculated density fields with various smoothing lengths to determine the
dependence of the properties of the density field on smoothing.  We
characterise the structure and evolution of the cosmic web by the PDF
of the density field, and by its moments, variance, skewness, and
kurtosis, using both variants of the definitions of these parameters,
mathematical and cosmological.  The information content in
the mathematical and cosmological variants of the PDF moments is identical,
but they characterise the properties of the cosmic web and its
evolution in a different way. To our knowledge, this is the first
study in which the PDF moments are investigated using both definition
methods, mathematical and cosmological, in a broad interval of
simulation redshifts and smoothing lengths.

The critical step in our study is the smoothing of the density field.
Smoothing enables us to select the populations of the cosmic web:
small-scale smoothing characterises the  web on the cluster-type scale, and  large-scale smoothing
describes the  web on the supercluster scale.
To characterise the evolution of populations of the cosmic
web, we use skewness and kurtosis evolutionary tracks and diagrams.
We calculated density fields using three smoothing recipes: $B_3$
spline, and cubic-cell and top-hat smoothing (described in Appendix A),
and found the respective moments. The comparison of the density fields and
moments for different smoothing recipes is given in Appendix B.
The sparsity of the density field meant that the results obtained with the $B_3$
spline are unusable, and it limited the usable range of smoothing scales for
cubic-cell and top-hat smoothing.

\subsection{Simulations of the evolution of the  cosmic web}  
 
To study the evolution of the parameters of the cosmological PDF, we
used a three-dimensional grid of input parameters: the box size of the 
simulation, $L_0$, the smoothing length, $R_t$, and the redshift, $z$.
The smoothing lengths of the original density fields from the numerical simulation
output have a cell size $L_0/N_{\mathrm{grid}}$, where $N_{\rm grid}$
is a parameter that typically ranges from 500 to 5000.  This we call
the smoothing rank zero. We used a smoothing recipe that increased the
smoothing length by a factor of 2.  We used this recipe successively four
to five times.  The use of regular sets of box lengths and smoothing
lengths yields simulation parameter sets with identical smoothing
lengths in units of $\Mpc$.  This allowed us to compare the properties of
the simulations with identical smoothing lengths, and in this way, we checked the
convergence of the results. An additional test was provided by the regularity
of the relations: PDF moments versus redshift $z$ or standard deviation
$\sigma$.  If deviations occur, their reasons can be found by
inspecting the respective PDFs.

{\scriptsize 
\begin{table}[ht] 
\caption{Simulation parameters  with the {\sc GADGET} code}
\centering
\begin{tabular}{lrrrrcc}
\hline  \hline
  Simulation   & $L_0$&$N_p$&$N_{\mathrm{real}}$&$m_p$&$R_{\mathrm{min}}$\\   
\hline  
(1)&(2)&(3)&(4) &(5)&(6)\\
  \hline  \\
  L256& 256&$512^3$&1&$0.993$&1\\
  L512& 512&$512^3$&1&$7.944$& 2\\
  L1024&1024&$512^3$&1&$63.55$&4  
 \label{Tab1}                         
\end{tabular} 
\tablefoot{Columns list (1) the name of the simulation; (2) the box size in
  $\Mpc$; (3) the number of particles; (4) the number of realisations; (5)
  the mass of a particle in units of $10^{10}h^{-1}M_\odot$; and (6)
  the minimum 
smoothing scale in units of $\Mpc$.} 
\end{table} 
}

We simulated the evolution of the cosmic web adopting a DM-only
$\Lambda$CDM model, using two sets of simulations. For the 
first set we used the {\sc GADGET}  code \citep{Springel:2005} with
three different box sizes $L_0=256,~512, \text{and}~1024~\Mpc$ with
$N_{\mathrm{grid}} = 512$.  The cosmological parameters for these
simulations are ($\Omega_m,\Omega_{\Lambda},\Omega_b,h,\sigma_8$)
=(0.28, 0.72, 0.044, 0.693, 0.84).

Initial conditions were generated using the COSMICS code by
Bertschinger (1995), assuming Gaussian fluctuations.  Simulations
started at redshift $z=30$ using the Zeldovich approximation. We
extracted density fields and particle coordinates for redshifts
$z=30,~10,~5,~3,~2,~1,~0.5,\text{and}~0$.  Table \ref{Tab1} shows the
simulation parameters. An analysis of the evolution of the power
spectra of 
these simulations is presented in Appendix A.4.  We extracted
the simulation output for eight epochs and used four smoothing scales at
each epoch, thus we had $3\times 8\times 4=96$ sets of
simulation parameters.

{\scriptsize 
\begin{table}[ht] 
\caption{Simulation parameters with the {\sc GLAM} code} 
\centering
\begin{tabular}{lrrrcccc}
\hline  \hline
  Simulation  &   $L_0$&$N_{\mathrm{grid}}$&$N_p$&$m_p$&$N_{\mathrm{real}}$&
   $R_{\mathrm{min}}$\\   
\hline  
(1)&(2)&(3)&(4) &(5)&(6)&(7)\\
\hline \\
 GLAM400&  400&3000&$1500^3$&0.16&8&0.53\\
 GLAM500& 500&5000& $2000^3$&0.13&24&0.40\\        
 GLAM1000& 1000&3000&$1500^3$&2.52&9&1.33\\
 GLAM1662& 1662&5000&$2000^3$&4.87&16&1.33\\
 GLAM2000&2000&3000&$1500^3$&20.2&9&2.66\\
 GLAM4000& 4000&4000&$2000^3$&68.2&20&4.00      
 \label{Tab2}                         
\end{tabular} 
\tablefoot{Columns list (1) the simulation
  name; (2)  the box size $L_0$ in units of $\Mpc$; (3) the number
  of grid elements; (4) the number of
  particles $N_p$; (5) the mass of a particle in units of $10^{10}h^{-1}M_\odot$;  (6)
  the number of realisations
  $N_{\mathrm{real}}$; (7)   the minimum
smoothing scale in units of $\Mpc$.} 
\end{table} 
}

The second set of simulations was made with the {\sc GLAM} code
\citep{Klypin:2018we}, which is a particle-mesh code that uses a large
grid size $N_{\rm grid}$. Because the {\sc GLAM} code is much faster than
{\sc GADGET}, we were able to use a much better mass resolution and
produced many realisations, which allowed us to reduce the effects of the cosmic
variance.  Different cosmological parameters were used for these
simulations: ($\Omega_m,\Omega_{\Lambda},\Omega_b,h,\sigma_8$)
=(0.307, 0.693, 0.044, 0.70, 0.828). The simulation parameters of the GLAM simulations are
given in Table~\ref{Tab2}.

Simulations started at  $z=100$ using the Zeldovich approximation.  For each
simulation, the smoothing had three ranks. The first rank had a cell four
times the size of the simulation grid cell
$L_0/N_{\mathrm{grid}}$. The second and third ranks had smoothing
radii two and four times larger, correspondingly.  Simulation outputs were stored at 6 to 17
redshifts at intervals of $0 \le z \le 20$.  For each set, we found the
density field $\delta$ in mean density units and calculated its one-point PDFs
and its moments, using smoothing recipes that we describe below.

\subsection{Dark matter density fields  and  moments of PDFs:
  Definitions and approximations}  

Each $N-$body simulation provides us with a population of DM particles  for a box of
size $L_0$ at redshift $z$. The density field was estimated using a filter of size $R_t$ with 
a total number of independent elements $N$. The density field was normalised to the average
matter density, providing us with the density contrast $\delta$,
\be
  \delta = \frac{\rho_{\rm DM}}{\Omega_m\rho_{\rm cr}}-1,
\ee
where $\Omega_m$ is the density parameter for the cosmological model and $\rho_{\rm cr}$ 
is the critical density of the Universe. The density distribution function $P(\delta)$ 
is defined as a normalised number of elements of the density field
with a density contrast in the range $[\delta,\delta+d\delta]$ 
\be
P(\delta) \equiv \frac{\Delta N}{N\Delta\delta}.
\ee
The second moment of $P(\delta)$ is the dispersion of the density field,
\be
\sigma^2 = \frac{1}{N} \sum_{j =1}^N{\delta_j^2}= \langle\delta^2\rangle.\ee
The third and fourth moments of the PDF are defined as the skewness and
kurtosis parameters $S$ and $K$,
\be
S =\frac{1}{N}\sum_{j=1}^N\left(\frac{\delta_j}{
    \sigma}\right)^3= \frac{\langle\delta^3\rangle}{\sigma^3},
%
\quad K = \frac{1}{N}\sum_{j=1}^N\left(\frac{\delta_j}{
        \sigma}\right)^4 -3= \frac{\langle\delta^4\rangle}{\sigma^4} -3.
\label{s4}
\ee
The additional term $-3$ in Eq.~(\ref{s4}) causes the value of $K=0$
for the Gaussian distribution. In statistics, this is called excess kurtosis.
These definitions are used in mathematical statistics and  in
many scientific fields. The skewness $S$ characterises the degree of
asymmetry of the distribution, while  the kurtosis $K$ measures the presence of
heavy tails and peaks in the distribution.  By definition, both are
dimensionless quantities \citep[e.g.][]{Kofman:1994aa,Bernardeau:1995aa}.

\begin{figure*}[ht] 
\centering 
\hspace{2mm}
\resizebox{0.95\textwidth}{!}{\includegraphics*{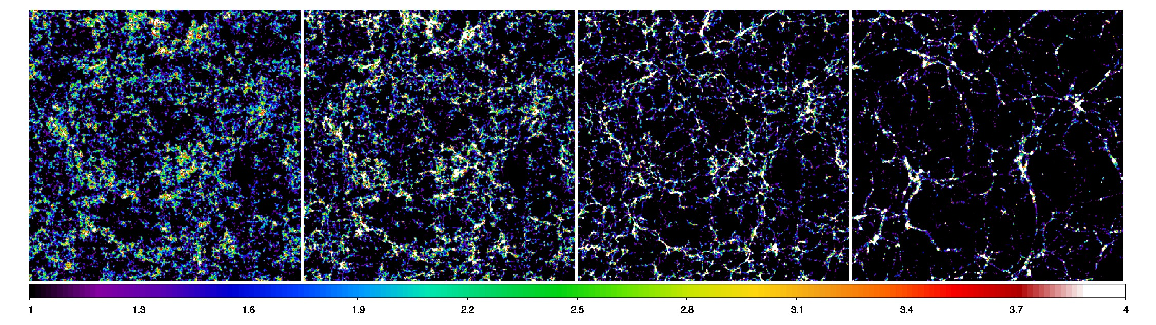}}
\hspace{2mm}
\resizebox{0.95\textwidth}{!}{\includegraphics*{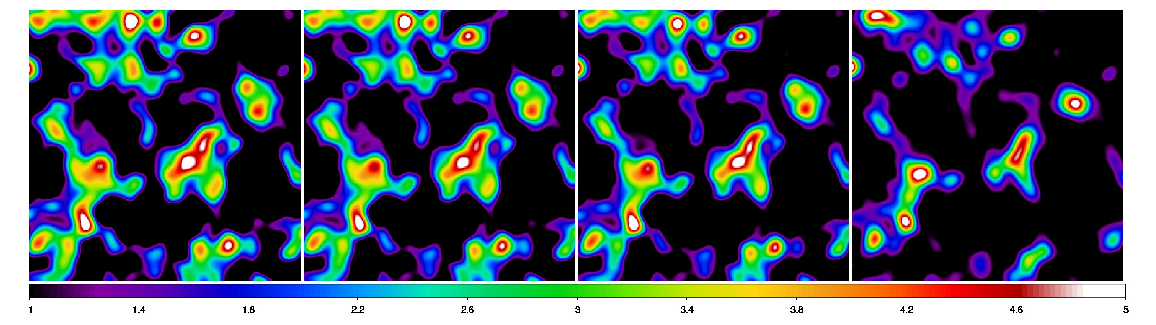}}
\caption{Density fields of simulation L256 without additional smoothing and with
  a smoothing length  $8~\Mpc$ are shown in the upper and bottom panels, respectively. The panels
  from left to right show fields for epochs $z=30$, $z=10$,  $z=3,$ and
  $z=0$,  presented in  slices of size
  $200\times200\times0.5~\Mpc$. Only over-density regions are shown
  with colour scales from left to right  $1-1.4,~1-2,~1-4,\text{and}~1-8$ in the upper
  panels and $1-1.08,~1-1.25,~1-1.8,\text{and}~1-5$ in the lower panels. 
   } 
\label{fig:Fig1} 
\end{figure*}

In the cosmological literature, another definition of the PDF moments is
used \citep[see][]{Peebles:1980aa,Bernardeau:2002aa,Szapudi:2009aa},
\be
S_p = \langle\delta^p\rangle/\sigma^{2(p-1)},\ee
where
\be
\langle\delta^p\rangle = \int_0^\infty\dd{\delta}P(\delta)\,\delta^p.
\label{moment}
\ee
These moments determine the $S_p$ parameters
\citep{Bernardeau:1995aa}. Specifically,
the third moment defines the skewness,
\be
S_3 = \langle\delta^3\rangle/\langle\delta^2\rangle^2,
\label{s3b}
\ee
and the fourth moment defines the  kurtosis,
\be
S_4 = \left(\langle\delta^4\rangle -
  3\langle\delta^2\rangle^2\right)/\langle\delta^2\rangle^3.
\label{s4b}
\ee
The second term in the last equation has the goal to obtain $S_4 =0$ for Gaussian
distribution.

By comparing mathematical  and cosmological definitions, it is easy to see
that
\be
S = S_3 \times \sigma,
\label{s3c}
\ee
and
\be
K= S_4 \times \sigma^2.
\label{s4c}
\ee
Equations~ (\ref{s3c}) and (\ref{s4c}) show that
mathematical skewness and kurtosis can be considered as power-law
functions of the standard deviation $\sigma$, where cosmological skewness
and kurtosis, $S_3$ and $S_4$, play the role of amplitude parameters
of mathematical $S$ and $K$.

We calculated smoothed density fields for the first set of models
L256, L512, and L1024 using three recipes, the $B_3$ spline, and
cell-cube and top-hat smoothing. Details of the calculations of
the density fields are explained in Appendix A.  For reasons explained
in Appendix A, we used the results obtained with the
cell-cube method for the core analysis.

We calculated for all density fields the mathematical skewness $S$ and
kurtosis $K$ using Eqs.~(\ref{s4}) and found
the cosmological skewness $S_3$ and kurtosis $S_4$ using Eqs.~(\ref{s3c}-\ref{s4c}). 
The variance $\sigma^2$, the skewness $S,$ and the
kurtosis $K$ were found with the {\tt moment} subroutine by
\citet{Press:1992aa}. This subroutine calculates the first four moments of the PDF. The
subroutine also calculates the standard deviation according to the
rule $Var= 1/(N-1) \sum\,\delta_j^2$. 

{ There are two ways to produce an approximation for the skewness $S_3$ and kurtosis $S_4$.
A dynamical theory can be used to predict the
PDF. Examples include the perturbation theory, the Zeldovich
approximation, or a spherical infall model. 
The perturbation theory   provides the following approximations
\citep{Juszkiewicz:1993aa,Bernardeau:1994aa,Kofman:1994aa}: 
\be
 S_3 = \frac{34}{7}+\gamma, \, S_4=\frac{60712}{1323}+\frac{62\gamma}{3}+\frac{7\gamma^2}{3}, \, 
 \gamma_1=\frac{d\log \sigma^2(R)}{d\log R},
 \label{eq:PT}
\ee
where $\gamma$ is the logarithmic slope of the dispersion $\sigma^2(R)$
with the filtering radius $R$. The parameter $\gamma$ is related to 
the effective slope of the power spectrum of perturbations at radius $R$ as $\gamma=-(n_{\rm eff}+3)$.
The approximations eqs.~(\ref{eq:PT}) are expected to work only for
small amplitudes of perturbations $\sigma\lesssim 0.1$. 

Another way to predict $S_3$ and $S_4$ is to use an
analytical form of the PDF and tune its parameters to make predictions. Examples
of this approach include the lognormal approximation
\citep[e.g.][]{Coles:1991vr, Lam:2008vy, Klypin:2018ac}, the skewed
lognormal distribution \citep{Shin:2017aa}, and the negative binomial
distribution \citep{Betancort-Rijo:2002ve}. The main assumption of these
analytical approximations is that the PDF only depends on the rms of
the density perturbations $\sigma$ and does not explicitly depend on
the redshift $z$. The widely used lognormal distribution is a good
example of this behaviour. It can be written as
\begin{eqnarray}
P(\rho)d\rho &=&\frac{1}{\sqrt{2\pi\sigma_1^2}}\exp\left( -\frac{\left[\ln\rho +\sigma_1^2/2\right]^2}{2\sigma_1^2} \right)\frac{d\rho}{\rho}, \\
\sigma_1^2&=&\ln(1+\sigma^2), \quad \rho =1 +\delta.
\end{eqnarray}
For the lognormal distribution, the skewness and kurtosis are
\be
S_3 = 3 + \sigma^2, \quad S_4 = 16 + 15\sigma^2 + 6\sigma^4 + \sigma^6.
\label{eq:LN}
\ee
}

{\subsection{Definitions: Cosmic web populations and their evolutionary tracks}

The presence of the cosmic web with clusters,  filaments,
superclusters, and  empty voids has been known for a long time. For early
observational evidence, see \citet{Gregory:1978},
\citet{Joeveer:1978pb}, \citet{Tarenghi:1978}, \citet{Tully:1978} and
\citet{de-Lapparent:1986}.
  For theoretical explanations see \citet{Zeldovich:1970}, \citet{Zeldovich:1978}, 
\citet{Zeldovich:1982kl}, \citet{Arnold:1982aa}, \citet{Bond:1996fv},
\citet{Bond:1996aa}, \citet{Pogosyan:2009aa}, and 
\citet{Cadiou:2020ab}. The basic constituent of the cosmic web is DM,
 which in the context of classical physics is a collisionless dust. 
   To select populations of  interest to cosmology, a smoothing of the density field
is needed. The smoothing scale determines the character of the populations.
To see the evolution of elements of the cosmic
web, we considered two main scales: clusters and superclusters.

The growth of structures on various scales during the evolution
of the cosmic web is shown in Figure~\ref{fig:Fig1}.  We plot here the density field of
simulation L256 at epochs $z=30$, $z=10$, $z=3,$ and $z=0$.  This
simulation has the highest resolution and allows showing the
evolution on galaxy up to supercluster scales better.  The upper panels show the
original L256 density fields with a cell size $L_0/N_{\mathrm{grid}}=0.5\,\Mpc$, 
 and the lower panels show the density fields smoothed with a length $R_t=8\,\Mpc$, using
the $B_3$ spline with a resolution $512^3$. This smoothing method and
scale are often used to determine galaxy superclusters,   see
\citet{Liivamagi:2012} and \citet{Einasto:2019fk}. We show only
over-densities where $D \equiv \delta + 1 \ge 1$.

\begin{figure*}[ht] 
\centering 
\hspace{2mm}
\resizebox{0.30\textwidth}{!}{\includegraphics*{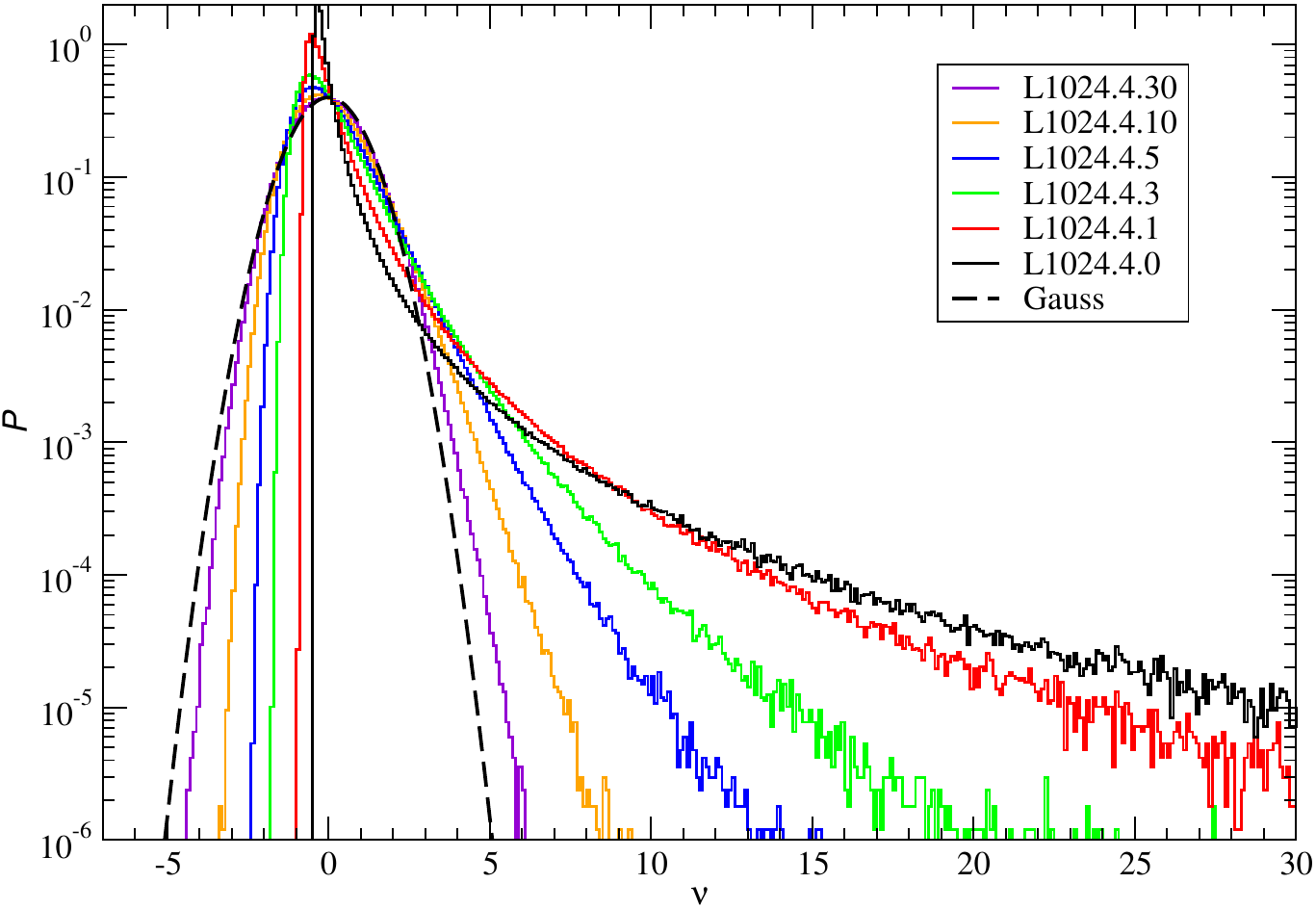}}
\hspace{2mm}
\resizebox{0.30\textwidth}{!}{\includegraphics*{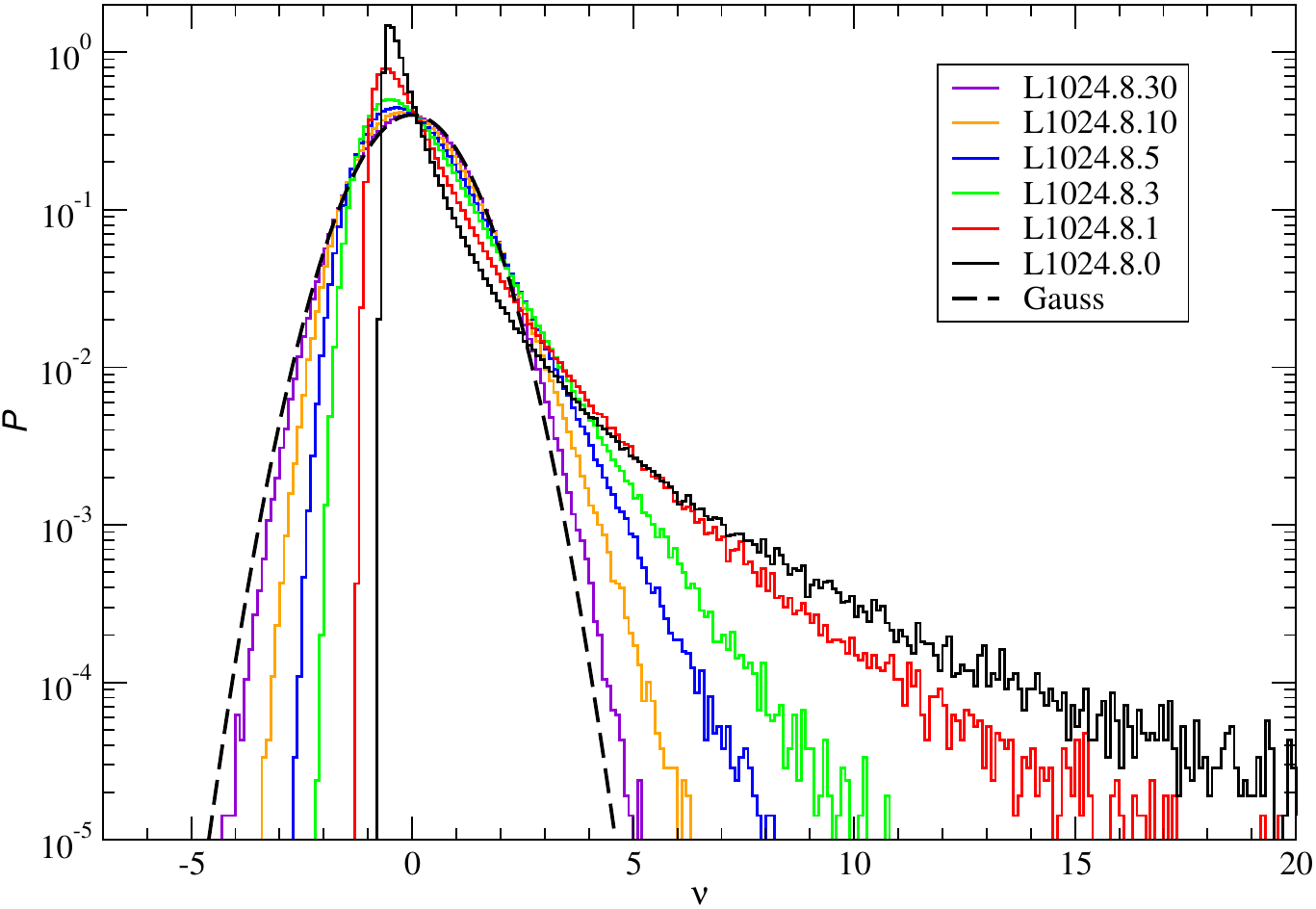}}
\hspace{2mm}
\resizebox{0.30\textwidth}{!}{\includegraphics*{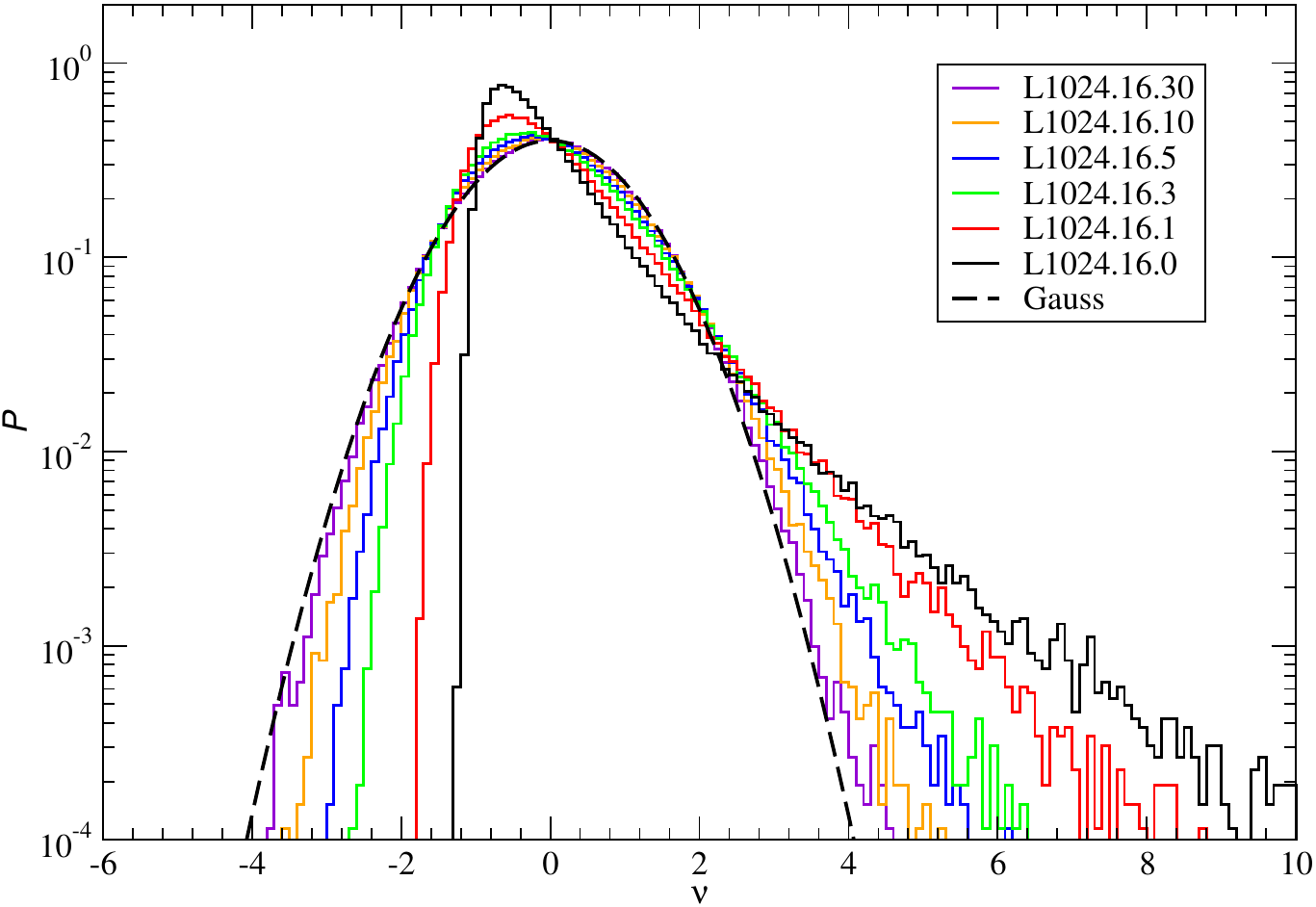}}
\caption{Density distribution functions $P(\delta)$ as functions of
  the DM density contrast $\delta$ normalised to the rms of density
  fluctuations $\nu = \delta/\sigma$ for simulation L1024 smoothed
  with the cubic-cell method. Panels from left to right show
  smoothing lengths $R_t=4,~8,~16~\Mpc$, indicated as the first index
  in the simulation name and the redshift coded in the second index.
  Dashed curves show the Gaussian distribution.}
\label{fig:Fig2} 
\end{figure*}

In the upper panels we show the evolution of small-scale elements of the
cosmic web,  galaxies and clusters of galaxies.  During the
evolution, they merge to form a sharp filamentary structure at the
present epoch.  This evolution is predicted by the theoretical models of 
\citet{Arnold:1982aa}, \citet{Bond:1996fv}, and others.
In the lower panels, the evolution of
supercluster-scale elements of the cosmic web is shown. Superclusters
alter their pattern very little during the evolution, 
only the amplitude of density fluctuations increases.

The density field method was used to select and study various
components of the cosmic web: clusters, filaments, superclusters,
voids, etc. These elements are individual objects, located in different
areas of the universe. Their volumes do not overlap, but in sum, these
components fill the whole universe.  There exists a large number of
various methods for investigating the structure and evolution of these
components of the universe. For recent studies, see proceedings of the
Zeldovich conference by \citet{VandeWeygaert:2016zt}.

Because of its integrated
nature, the PDF does not allow selecting individual components of the
cosmic web. The PDF of the density field and its moments are
integrated quantities that characterise properties of the whole web.  
Objects of various compactness of the cosmic web can be
highlighted using a smoothing of the density field with different
scales. Examples of various smoothing scales were shown in
Figure~\ref{fig:Fig1}.  Smoothing with small lengths, $R_t \le 1\,\Mpc$,
highlights the whole cosmic web in the volume under study on scales of
haloes and subhaloes of ordinary galaxies and poor clusters.  Medium
smoothing lengths, $R_t =2,~4\,\Mpc$, are suited to highlighting the
cosmic web on scales of rich clusters of galaxies and the central regions
of superclusters.  A large smoothing with $R_t =8 - 16\,\Mpc$ highlights
the cosmic web on the supercluster scale.

Instead of ``cosmic web at smoothing scale $R_t$'' , we use the term 
``populations of the cosmic web'' according to the smoothing length that is applied to
calculate the density field.  Populations cover the whole cosmic
web, they characterise the web on the selected smoothing scale. The
smoothing scale is a physical parameter that allows highlighting the
cosmic web at the scale  of interest. Using various smoothing lengths, we can study the 
hierarchy of structures in the cosmic web.

To quantitatively describe the evolution of the populations of the cosmic
web, we used the skewness $S$ and kurtosis $K$ ($S_3$ and $S_4$) as
functions of the age of the universe, measured by redshift $z$, or as
functions of the rms of the density field, $\sigma$.  These functions
depend on the power spectrum index $n$ and on other cosmological
parameters of the model ($\Omega_m$, $\Omega_\Lambda$). The simulation
epoch, $z,$ and the smoothing length, $R_t$, are parameters.  Every
simulation of the evolution with fixed values of the parameters $z$ and
$R_t$ yields a dot in the evolutionary diagrams.

We call the graphs in which lines join simulations with a given smoothing
length $R_t$ at various $z$ the ``evolutionary tracks'' of cosmic web
populations, and the graphs in which lines join simulations with various
smoothing length $R_t$ at a given epoch $z$ the ``evolutionary
diagrams''.  This follows the analogy with stellar evolution tracks and
Hertzsprung-Russel diagrams. Evolutionary tracks show  evolutionary
trajectories  of
populations of the cosmic web at various characteristic scales in the
$S(z)$, $K(z)$, $S(\sigma)$, $K(\sigma),$ and $S_{3,4}(z)$,  $S_{3,4}(\sigma)$ plots.
Evolutionary diagrams show where characteristic populations
of the cosmic web are situated in  these diagrams at various
epochs.  Evolutionary tracks and diagrams are based on identical data,
only the data points are joined differently by the lines.

Evolutionary tracks and diagrams display the
evolution of the asymmetry of the cosmic web in a simple way.  Each
plot shows at a glance the growth of the asymmetry 
of the whole web on different scales.

\begin{figure*}[ht] 
\centering 
\hspace{2mm}
\resizebox{0.75\textwidth}{!}{\includegraphics*{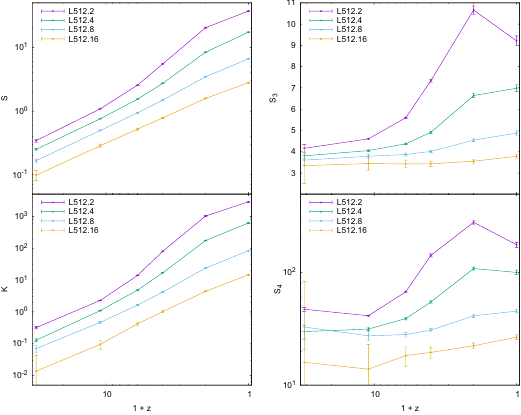}}
\caption{Dependence of the PDF moments  on the
  cosmic epoch $z$ for simulation L512. The {\em left} top and bottom panels show the
  dependence for the skewness $S(z)$ and the kurtosis $K(z)$,
  respectively,  and the {\em right} panels show their cosmological equivalents
  $S_3(z)$ and $S_4(z)$.  The index in the simulation name is the smoothing length
  in $\Mpc$. }
\label{fig:Fig3} 
\end{figure*} 

\section{Analysis}

In this section we describe the evolution of the cosmic web in the
$\Lambda$CDM model through its PDF moments.  Next, we investigate the
evolution of the skewness $S$ and the kurtosis $K$.  Thereafter, we
analyse skewness and kurtosis as functions of rms of the density field
$\sigma$ and their change with cosmic epoch and smoothing length.
These relations are described first for mathematical and then for
cosmological moments.  In the first subsections, we mostly use results
obtained with GADGET simulations.  In the last subsection, we present
results obtained with both simulation series.

\subsection{Evolution of the PDF of $\Lambda$CDM simulations}

Figure~\ref{fig:Fig2} shows the evolution of the PDF using the cubic-cell
smoothing window.  We used as argument the reduced density
$\nu=\delta/\sigma$.  This presentation is useful to show how the
density distributions of our simulations can be represented by a 
Gaussian distribution.  We show the density fields of simulation L1024
using three smoothing lengths $R_t=4,~8,~\text{and }16\,\Mpc$ from
left to right. The  PDFs of simulations L512 and L256 are very
similar.  The colour-coding indicates the evolutionary epoch,
$z=30,~10,~5,~3,~1,\text{and}~0$. 

The PDFs of our density fields, obtained with cubic-cell  
smoothing, are very similar to the PDFs found in previous studies for
epochs $z \le 3$; for an example, see \citet{Shin:2017aa}.  All
$P(\nu)$ curves in low-density regions, $\nu \le 0$, lie below the
Gaussian curve and are in high-density regions, $\nu \ge 5$, above
the Gaussian PDF. This behaviour is expected for PDFs with positive skewness.

Two conclusions are evident from this figure: (i) PDFs are
asymmetric in the sense that high-density regions extend much farther
than low-density regions; and (ii) smoothing has a dominant role in
determining the width of the PDF distribution.  Both the asymmetry and
the importance of smoothing role of PDFs have been known for a long time; for early work, see
\citet{Bernardeau:1994aa}, \citet{Kofman:1994aa}, and
\citet{Bernardeau:1995aa}.  The dependence of PDFs on smoothing length
was recently studied by \citet{Shin:2017aa} and \citet{Klypin:2018ac}.
In our study, we see the growth of the asymmetry in a very broad
redshift interval, from $z = 30$ to $z = 0$.

\subsection{Evolution of the variance, skewness and kurtosis with
  cosmic epoch $z$}

Figure~\ref{fig:Fig3} presents the dependence of the PDF moments of
simulation L512 on the cosmic epoch $z$. In the left panels the dependence
is shown for the skewness $S(z)$ and the kurtosis $K(z)$, and in the right
panels for the respective cosmological functions $S_3(z)$ and
$S_4(z)$. Coloured lines joining symbols are the evolutionary tracks of
populations of various richness, identified by the smoothing lengths. The
horizontal axes are inverted to show the evolution from left to right
as in the following figures.  If we join points at given epochs $z$, we
obtain evolutionary diagrams. In this representation, they are vertical
lines.{ The error bars shown in this and the following figures were
calculated using the recipes described in Appendix~\ref{app.error}.}

\begin{figure}[ht] 
\centering 
\hspace{2mm}
\resizebox{0.45\textwidth}{!}{\includegraphics*{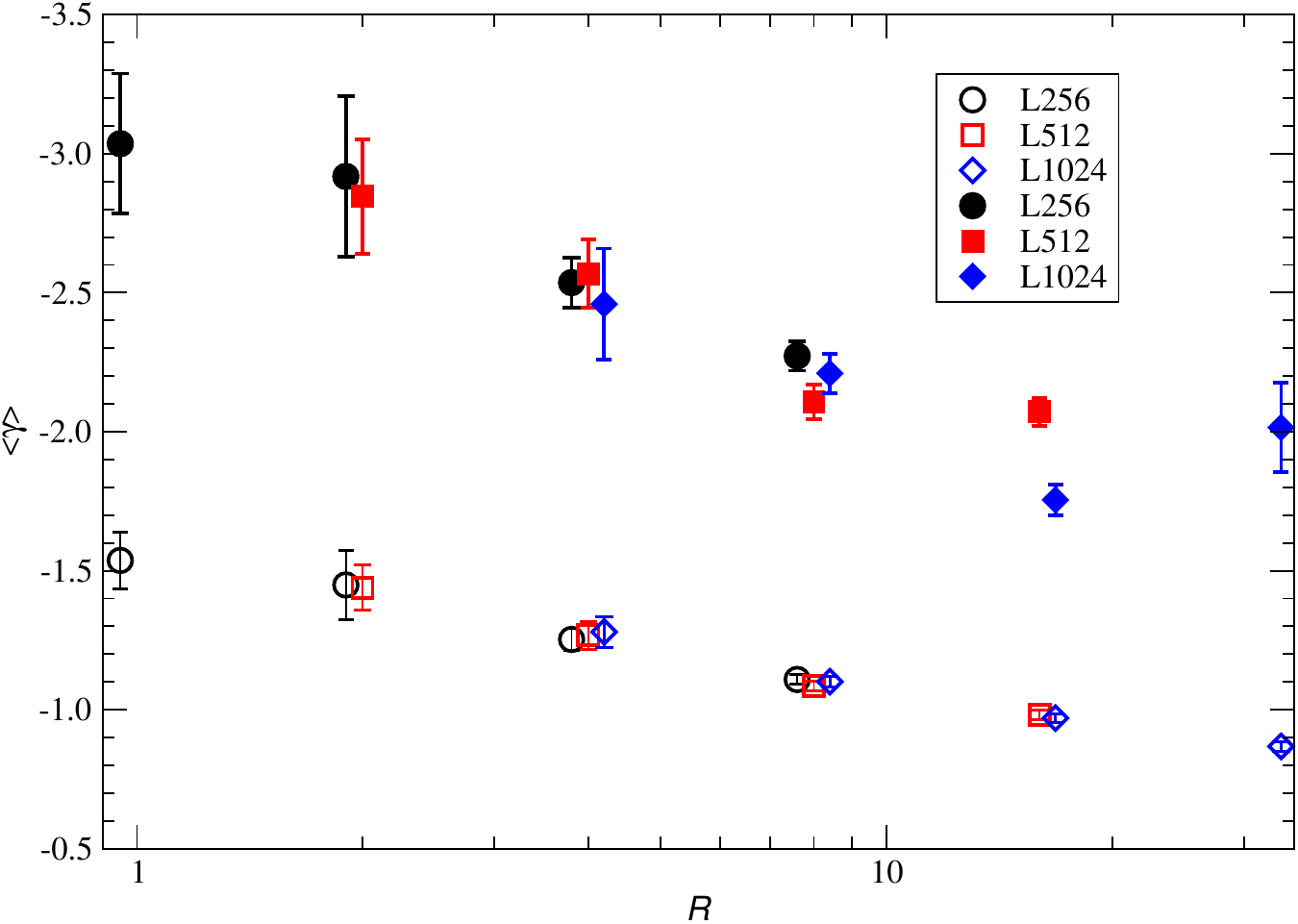}}
\caption{{ Mean logarithmic gradients for the redshift range
$0 \le z \le 30$}  of the skewness,
  $\langle\gamma_S\rangle$, and the kurtosis, $\langle\gamma_K\rangle$, as
  functions of the smoothing length, $R_t$. Open symbols show
the  skewness $S$, and filled symbols show the kurtosis $K$. For clarity, the
  symbols for simulations L256 and L1024 are slightly shifted in $R_t$
  coordinate. }
\label{fig:Fig4} 
\end{figure} 

\begin{figure*}[ht] 
\centering 
\hspace{2mm}
\includegraphics[width=0.75\textwidth]{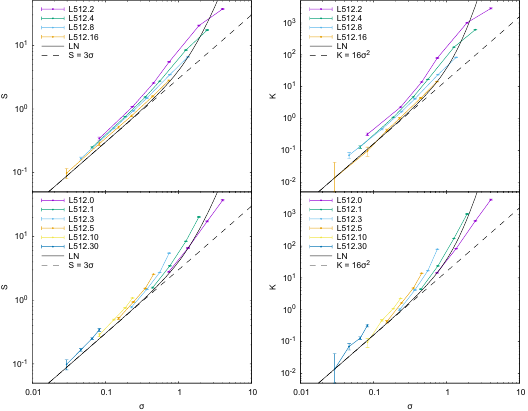}
\caption{ { {\em Top panels} show evolutionary tracks. {\em Bottom
      panels} show evolutionary diagrams. {\em Left panels} show the skewness
    $S$. {\em Right panels}  show the kurtosis $K$. In the evolutionary tracks, the
    symbols along the tracks from left to right are for redshifts
    $z=30,~10,~5,~3,~1,\text{ and}~0$.  In the evolutionary diagrams, the symbols from top
    to bottom correspond to smoothing lengths $2,~4,~8,\text{and } 16~\Mpc$.
    The solid bold black curves show the skewness $S(\sigma)$ and kurtosis
    $K(\sigma)$ according to the log-normal  distribution. The dashed
    curves show $S=3\sigma$ and $K=16\sigma^2$.}  }
\label{fig:Fig5} 
\end{figure*} 

The evolution of the skewness $S(z)$ and of the kurtosis $K(z)$ are
dominated by the increase in variance $\sigma^2(z)$ with time.
The growth of the skewness $S(z)$ is approximately proportional to the
growth of $\sigma(z)$, and the growth of the kurtosis $K(z)$ is
proportional to the growth of the variance $\sigma^2(z)$. During the
evolution from $z=30$ to $z=0,$ the amplitude of the skewness $S(z)$
increases about 30 times and that of the kurtosis $K(z)$ increases by a factor
of one thousand.  The other important aspect is the dependence of
the amplitudes of the skewness and kurtosis curves on the smoothing length
$R_t$.  At the present epoch, the value of the skewness $S(0)$ is for
the smoothing length $R_t=2\,\Mpc$, ten times higher than for smoothing
length $R_t=16\,\Mpc$; the difference in kurtosis $K(0)$ is two orders
of magnitudes.

The rate of the
evolution of moments can be characterised by the logarithmic gradient,
\be
\gamma_S(z) =  \frac{d\log\,S(z)}{d\log\,(1+z)}
\label{eq.gamma}
,\ee
for the skewness $S(z)$ and a similar relation for the kurtosis $K(z)$.
{ Figure~\ref{fig:Fig4} shows the mean logarithmic gradients for
  the redshift range $0 \le z \le 30$ of the skewness and kurtosis,
  $\langle\gamma_S\rangle$ and $\langle\gamma_K\rangle$, as functions
  of the smoothing length, $R_t$.}  Figures~\ref{fig:Fig3} and
 \ref{fig:Fig4} show that the negative gradients of the skewness,
$\gamma_S$, and of the kurtosis, $\gamma_K$, change with epoch and
smoothing length $R_t$.

The rapid change in skewness $S(z)$ and kurtosis $K(z)$ with
cosmic epoch $z$ is eliminated when their cosmological
equivalents, $S_3(z)$ and $S_4(z)$, are used, which we show in the right panels of
Figure~\ref{fig:Fig3}.  Here the dependence of the evolution of the
skewness $S_3(z)$ and of the kurtosis $S_4(z)$ on the smoothing length
$R_t$ is very clear.

We note that the cosmic epoch $z$ and the smoothing length $R_t$
uniquely determine the position of the model universe in $S(z)$, $K(z)$,
$S_3(z),$ and $S_4(z)$ functions and vice versa: any fixed value of
these functions uniquely determines the $z$ and $R_t$ parameters of the models (for
identical  cosmological parameters).

\begin{figure*}[ht] 
\centering 
\includegraphics[width=0.75\textwidth]{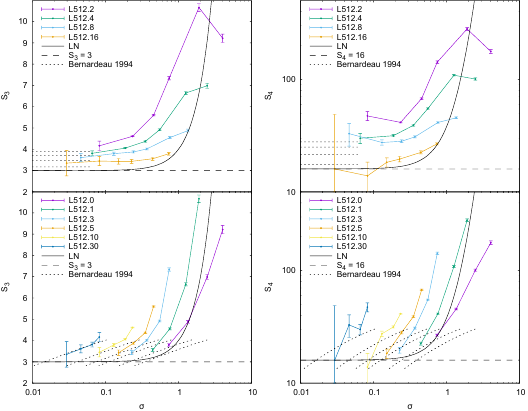}
\caption{  {  {\em Top panels} show evolutionary tracks. {\em Bottom panels}
  show evolutionary diagrams. {\em Left panels} show the skewness $S_3$. {\em
    Right panels} show the kurtosis $S_4$. In the evolutionary tracks, the symbols
  along the tracks from the left are for redshifts $z=30,~10,~5,~3,~1,\text{ and}~0$.  In
the  evolutionary diagrams, the symbols from top to bottom correspond to
  smoothing lengths $2,~4,~8,\text{ and } 16~\Mpc$.  The solid black curves show the
  cosmological skewness $S_3$ and kurtosis $S_4$ according to the log-normal
   distribution. The dotted curves show the predictions of the
  perturbation theory  Eqs.~(\ref{eq:PT}).}  } %
\label{fig:Fig6} 
\end{figure*} 

\subsection{Evolutionary tracks and diagrams of cosmic
  web populations in $S(\sigma)$ and  $K(\sigma)$ }

In Figure~\ref{fig:Fig3} the evolutionary routes of populations are shown
using as argument the epoch $z$.  Another possibility to present
the evolutionary routes is to use as argument   the
density field $\sigma$ instead of $z$.    The relations
$S(\sigma)$ and $K(\sigma)$ were first investigated by
\citet{Kofman:1994aa}. We show these relations in Figure~\ref{fig:Fig5}
for simulation L512. The  left panels show the skewness $S(\sigma)$ and  right
panels the kurtosis  $K(\sigma)$. In the top panels, the evolutionary tracks are presented,
and curves join various populations. In  the bottom panels, the evolution
diagrams are shown. Here,  simulations of identical age are
connected by coloured curves that link the four smoothing lengths
$2,4,8,\text{and }16~\Mpc$. Error bars are also marked.  Other simulations
yield similar pictures.  Because of the differences in 
resolution, data for simulation L256 are slightly shifted toward
higher $\sigma$ and data for the L1024 simulation toward lower
$\sigma$.

The populations have different values of $\sigma$. In this
representation, populations of the same age but different smoothing
length are therefore shifted along the $\sigma$ coordinate.  It is remarkable
that the shift in $\sigma$ causes evolutionary tracks of different
smoothing lengths to almost coincide, as shown in
Figure~\ref{fig:Fig5}.  This coincidence led \citet{Kofman:1994aa} to
the conclusion that a range of $\sigma$ values could be obtained
either by analysing the system at different epochs $z$ or by using
different smoothing lengths $R_t$. As we show below, the dependence of
the skewness and kurtosis on $z$ and $R_t$ is different.

Figure~\ref{fig:Fig5} shows that the overall mean evolution of the
skewness $S(\sigma)$ and the kurtosis $K(\sigma)$ is proportional to
the first and second power of $\sigma$, in accordance with the definition
Eqs.  (\ref{s3c}) and (\ref{s4c}).  The figure also shows that the
evolution on various scales is different.  We discuss these
differences in more detail in the next subsection.

\begin{figure*}[ht] 
\centering 
\includegraphics[width=0.45\textwidth]{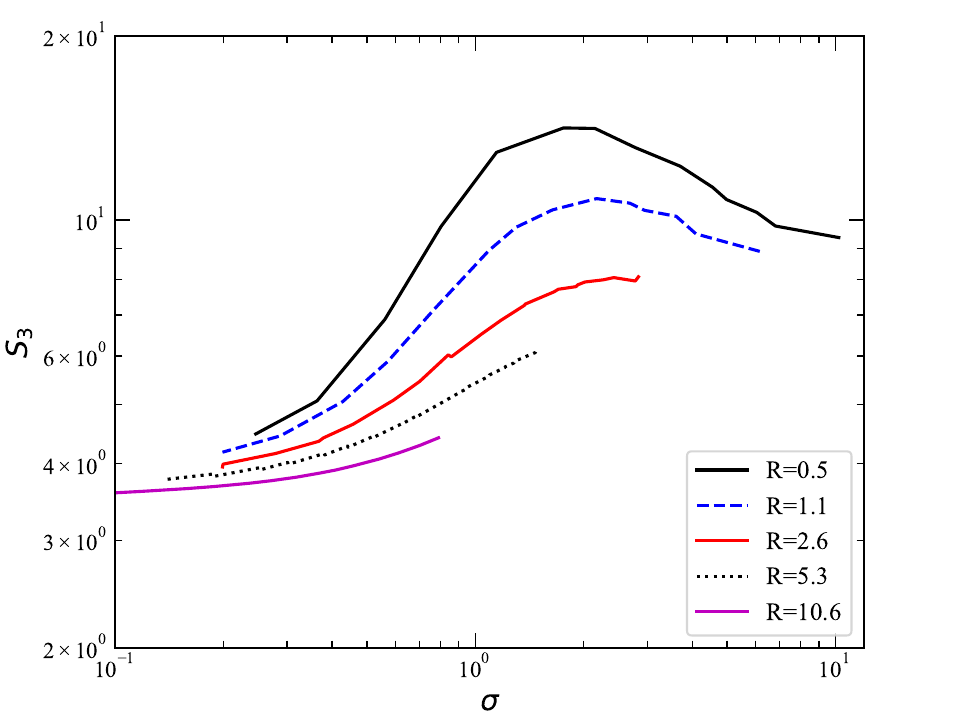}
\includegraphics[width=0.45\textwidth]{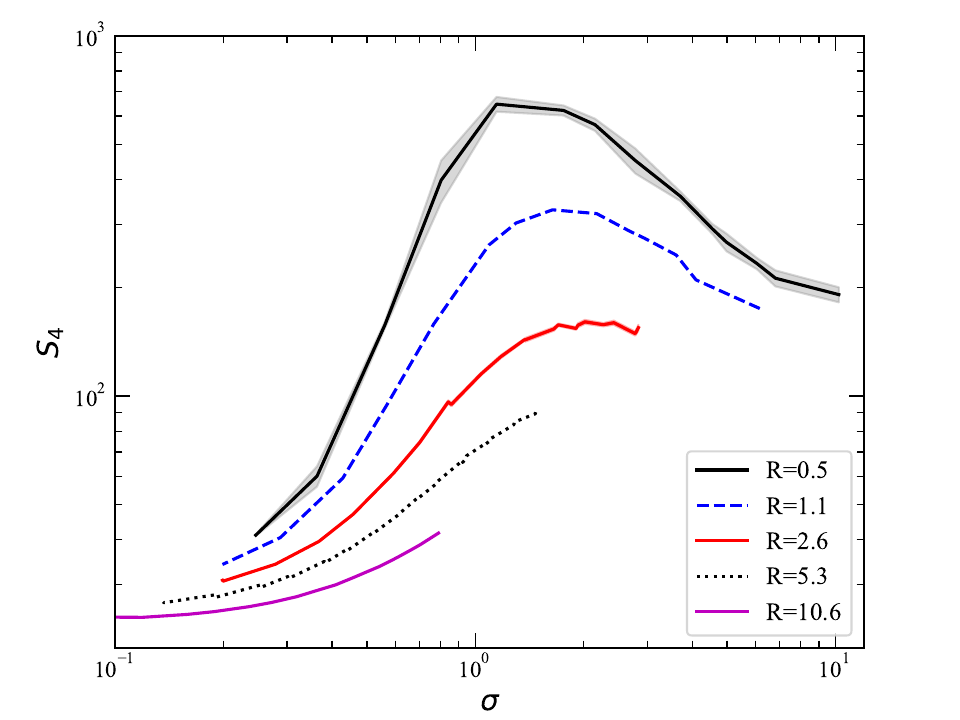}
\includegraphics[width=0.45\textwidth]{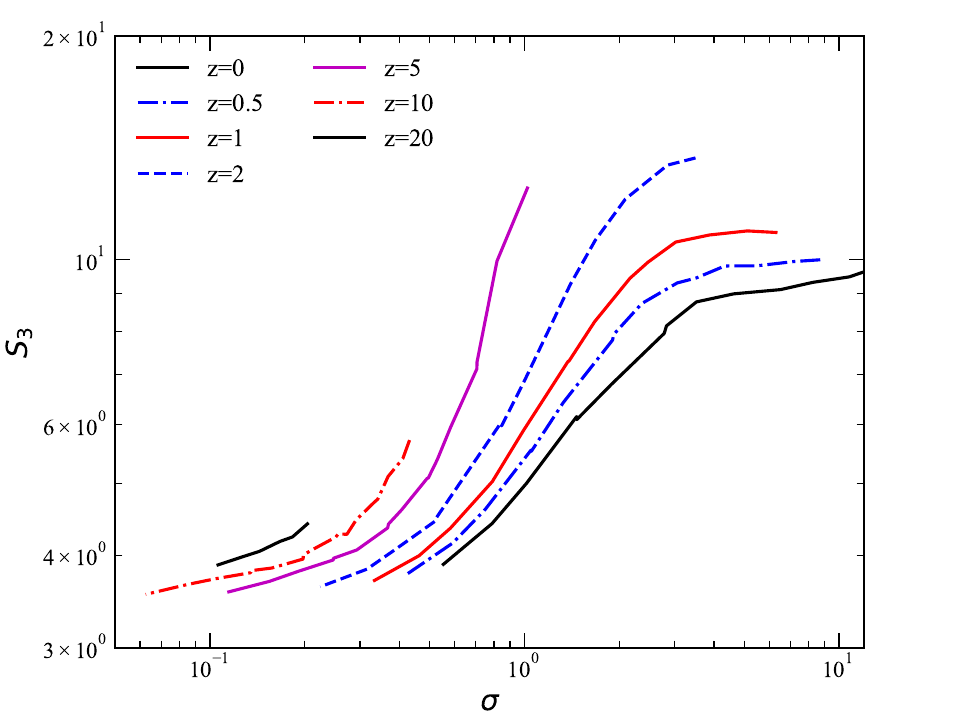}
\includegraphics[width=0.45\textwidth]{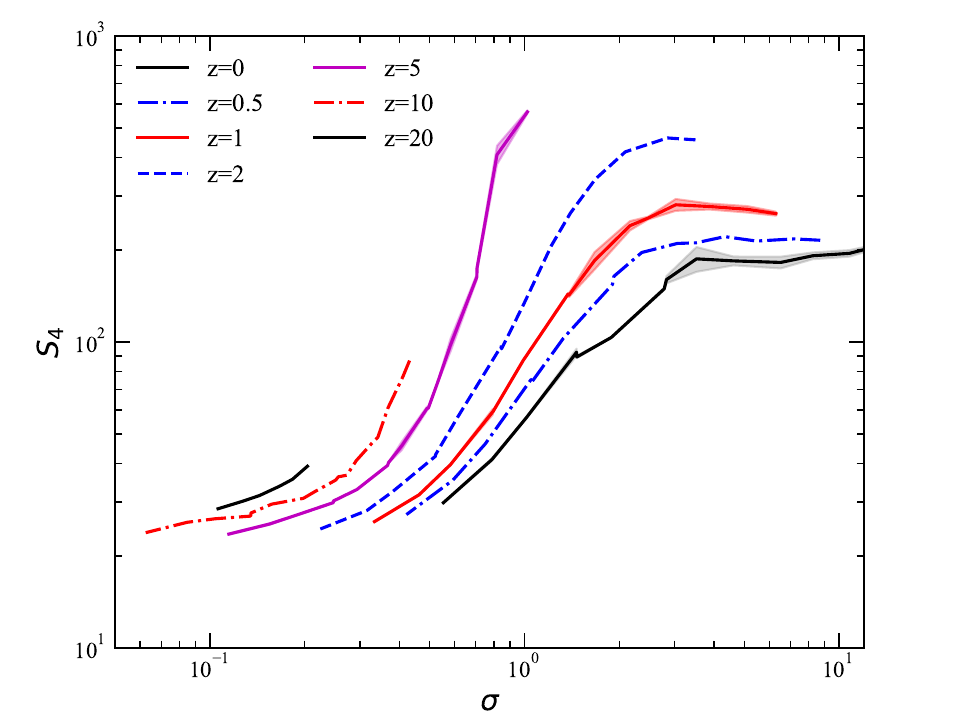}
\caption{  {\em Top panels} show evolutionary tracks. {\em Bottom panels}
  show evolutionary diagrams. {\em Left panels} show the skewness $S_3$. {\em
    Right panels} show the kurtosis $S_4$. 
  The shaded areas correspond to the $1\sigma$ statistical uncertainties.}
\label{fig:Fig7} 
\end{figure*}

Figure~\ref{fig:Fig5} also shows $S(\sigma)=S_3\,\sigma$ and
$K(\sigma)=S_4\,\sigma^2$ curves for the log-normal PDF distribution.  At small rms of the density field, 
$\sigma \le 0.5$, the log-normal distributions of $S(\sigma)$ and
$K(\sigma)$ are power laws with indices $1$ and $2$, respectively.
For larger $\sigma,$ the log-normal law bends upwards to imitate the
dependence of $S(\sigma)$ and $K(\sigma)$ on the smoothing length.

\subsection{ Evolutionary tracks and diagrams of cosmic web populations
  in $S_3(\sigma)$ and  $S_4(\sigma)$ }

The evolution of the skewness $S(\sigma)$ and kurtosis $K(\sigma)$ is
dominated by changes in the rms of the
density field $\sigma$. A much more
compact presentation of the evolution is possible when we use
cosmological parameters according to the definitions Eqs. ~ (\ref{s3c})
and (\ref{s4c}), $S_3(\sigma) = S(\sigma)/\sigma$ and
$S_4(\sigma)=K(\sigma)/\sigma^2$.

This version of evolutionary tracks and diagrams is presented in
Figure~\ref{fig:Fig6}. It is based on data for simulation L512.  { 
  As explained in Section 2, in evolutionary tracks (top panels), the
  coloured curves join the $S_3(\sigma)$ and $S_4(\sigma)$ values for
  various smoothing lengths $R_t$.}  Moving along the tracks from
left to right, the asymmetry and flatness parameters of the
cosmic web at given smoothing lengths change with redshift.  {
  The populations selected with $R_t \le 2~\Mpc$ and shown by cyan
  curves} have maxima $S_3\approx 11$ at $z \approx 2$, and decrease
in $S_3$ for later epochs, $z \le 2$.  The populations selected with a
smoothing length $R_t= 4~\Mpc$ and shown by green curves reach
amplitudes $S_3 \approx 7$ at the present epoch.  { The populations
  selected 
  with a smoothing length $R_t=8~\Mpc$ and shown by light blue curves
  have a moderate increase in $S_3(\sigma)$.}  A similar increase
was found by \citet{Shin:2017aa} for epochs $z \le 4$, see their
Figure~4. The authors used top-hat smoothing with $R_t=10~\Mpc$.
The populations selected with a smoothing length $R_t=16~\Mpc$ have
approximately constant $S_3(\sigma)$ levels during 
the evolution; see the
orange curve in the top left panel of Figure~\ref{fig:Fig6}.

The evolutionary tracks of the kurtosis  $S_4(\sigma)$ are shown in the
top right panel  of Figure~\ref{fig:Fig6}.  The general shape of tracks is 
similar, only the growth of $S_4(\sigma)$ for small smoothing lengths and
late epochs is much stronger. The  $S_4(\sigma)$ evolutionary tracks are more
affected by errors, both sampling errors, shown as error bars in
Figure~\ref{fig:Fig6}, and possible systematic errors, discussed in
Appendix B. When we take these possible errors into account, the
increase in $S_3(\sigma)$ and $S_4(\sigma)$ 
with decreasing smoothing length is a general property of the
evolution.

The bottom panels in Figure~\ref{fig:Fig6} show evolutionary diagrams of
the populations for simulation L512: curves joining symbols connect
simulations of identical redshift $z$. As in Figure~\ref{fig:Fig5}, for each
epoch, the symbols from the top down correspond to smoothing lengths 
$R_t=2,~4,~8, ~\text{and }16~\Mpc$.  The evolutionary 
diagrams for different ages are well separated from each other and
are located at approximately similar mutual distances along the
$\sigma-$axis.  This conclusion is valid for both $S_3(\sigma)$ and
$S_4(\sigma)$.  

{ We show in all panels the $S_3(\sigma)$ and
  $S_4(\sigma)$ functions as predicted by the perturbation theory,
  Eqs.~(\ref{eq:PT}), for the same set of smoothing lengths and
  redshifts.  The comparison shows that for redshift $z=30$ and large
  smoothing lengths, the PT is in fairly good agreement with the results of
  numerical simulations.  For a lower redshift and smaller smoothing lengths, the
  PT is not able to reproduce the $S_3(\sigma)$ and $S_4(\sigma)$
  functions found in simulations. Differences increase with the
  decrease in smoothing length.}

Figure~\ref{fig:Fig7} shows the results of the
GLAM simulations for the skewness $S_3(\sigma)$ and kurtosis
$S_4(\sigma)$. Here we combined
data from  six GLAM  simulations, all with many realisations, therefore the shot noise
is much smaller. Data for different boxes agree quite well within
$2 - 3$~\%.  In the top panels, the coloured curves join simulations with
identical smoothing lengths, and in the bottom panels, they join
simulations of identical redshift, that is, we have evolutionary
tracks and diagrams, respectively.

The evolutionary tracks of populations are shown in great detail,
especially for populations with small smoothing lengths. At the lowest
smoothing length, the peaks of $S_3(\sigma)$ and $S_4(\sigma)$ are at
redshift $z=5$, followed by a slow decrease at lower
redshifts.  The evolution is shown for five smoothing lengths from
$R_t=0.5~\Mpc$ to $R_t=10.6~\Mpc$.  Moving along tracks from left to
right,  $S_3(\sigma)$ and $S_4(\sigma)$ change during the
evolution from $z=10$ to $z=0$. In this figure, the points for various
redshifts are not marked.

The evolutionary diagrams were calculated for seven redshifts, starting
from $z=20$.  The lower tips of the curves in most cases correspond to
the smoothing length $R_t=10.6~\Mpc$, and the upper tips show the lowest smoothing
length $R_t=0.5~\Mpc$.  Figure~\ref{fig:Fig7} shows that the evolutionary
diagrams have a more complex structure than expected from simulations
with lower resolution. For smaller smoothing lengths, $R_t \le 2~\Mpc$, and recent
epochs, $z \le 1$, the diagrams reach constant levels.  All curves
for changing $z$ and $R_t$ yield monotonic ladders without crossing
each other.  The GLAM simulations confirmed all basic findings
from the GADGET simulations with high confidence, and
they suggested some important details that were not observed in simulations with low-mass resolution.

\begin{figure}[ht] 
\centering 
\hspace{2mm}
\resizebox{0.45\textwidth}{!}{\includegraphics*{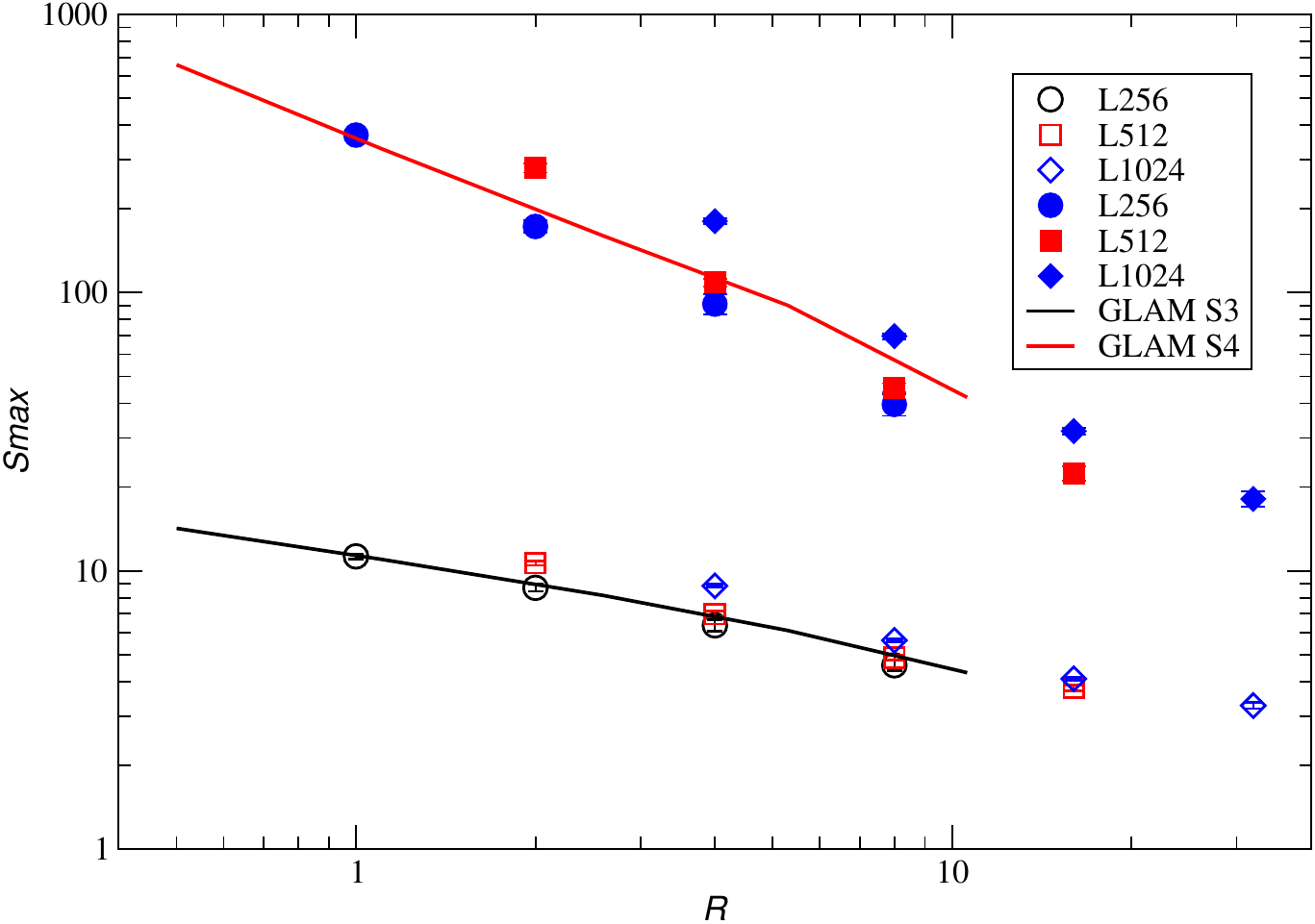}}
\caption{Maxima of the skewness $S_3$ and kurtosis $S_4$
  evolutionary tracks as functions of the smoothing length $R_t$. Open
  symbols show the skewness $S_3$ and filled symbols show the kurtosis
  $S_4$ for GADGET simulations of various cube lengths $L_0$. The
  black and red curves show the maxima of the $S_3$ and $S_4$ curves of the GLAM simulations.
}%
\label{fig:Fig8} 
\end{figure}

{ The comparison of Figures \ref{fig:Fig6} and \ref{fig:Fig7} shows
  that the evolutionary tracks found with the   GADGET and GLAM
  simulations  are qualitatively very similar.  Both start at high
  redshifts (small 
  $\sigma$) at
  levels $S_3 \approx 3.5$ and $S_4 \approx 25$, and have maxima of the
  $S_3(\sigma)$ and $S_4(\sigma)$ curves for different smoothing
  lengths at similar $\sigma$ and redshift $z$ values.  In
  Fig.~\ref{fig:Fig8} we show the maxima of the skewness $S_3$ and kurtosis $S_4$
  evolutionary tracks as functions of the smoothing length
  $R_t$. The redshifts of the maxima depend on the smoothing length $R_t$.  For
  a small smoothing length $R_t=0.5~\Mpc,$ the maxima are located at redshift
  $z \approx 5$, and for $R_t=1~\Mpc, $ they lie at $z\approx 1$. For greater
  smoothing lengths $R_t \ge 2.6,$ the maximum is not reached, see
the  upper panels of Fig.~\ref{fig:Fig7}. In these cases, we accepted as
  maximum the $S_3$ and $S_4$ value at redshift $z=0$. We ignore in
  Fig.~\ref{fig:Fig8} the redshift dependence of the maxima.

  Figure~\ref{fig:Fig8} shows that an almost linear
  relationship exists between maxima and smoothing length in the log-log
  presentation.  There is a scatter of  the maxima that is found with GADGET
  simulations for different cube lengths, which is larger than
  expected from the sampling errors. However, the overall trend with $R_t$
  is similar for all simulations, thus the scatter is likely caused by
  difficulties of determination of moments, as discussed in Appendix
  B.  The curves $S_3(R_t)$ and $S_4(R_t)$ for the GLAM simulations are
  located within the range expected from the GADGET simulations. Thus we
  see that GADGET and GLAM simulations yield very similar results for
  PDF moments in quantitative terms as well.

  The presence of the maxima in the evolutionary tracks shows the change in
  the  rate of the growth of the asymmetry of the cosmic web, measured by
   the logarithmic gradients $\gamma_S(z)$ and $\gamma_K(z)$, see
   Eq.~(\ref{eq.gamma}).  At the maxima of $S_3(z)$ and $S_4(z),$
   the gradients $\gamma_S(z)$ and $\gamma_K(z)$ change, that is, the rate of
   the growth of the asymmetry slows. On smaller scales, highlighted
   by small $R_t$, the change occurs at higher redshifts. 
  }

\section{Discussion}

In this section we discuss the growth of the density fluctuations and the
evolution of the particle densities using theoretical models.
We then compare our results with earlier results.  Finally, we
discuss the cosmological interpretation of our data.

\subsection{Comparison with theoretical models}

An important conclusion from our data is that the dependences of the PDF
moments on the evolutionary epoch $z$ and on the smoothing length $R_t$ are
very different.  Figures~\ref{fig:Fig6} and \ref{fig:Fig7} clearly
demonstrate that the rms of the density fluctuations $\sigma$ does not
determine 
the moments of the PDF in a unique way.  At a fixed redshift, that is,  in
the evolutionary diagrams,  the $S_3$ and
$S_4$ curves increase with $\sigma$ at small $\sigma\lesssim 1$,  then
they flatten out and stay constant at $\sigma\gtrsim 2$. The behaviour is
different for fixed $R_t$, that is, in evolutionary tracks.  At small
$\sigma,$ the curves turn upwards, but 
then they reach a maximum and start to decline.  Regardless
of the selection (constant $z$ or constant $R_t$), the curves show a
complex behaviour.  For example, the position and amplitude of the maximum
of $S_3$ change with $R_t$ if $R_t$ is fixed.  For fixed $z,$ the
amplitude and the redshift $z$ of the plateau depend on the redshift.

The results presented in Figures~\ref{fig:Fig6} and \ref{fig:Fig7} show 
interesting and somewhat counter-intuitive features: at the same rms  of the fluctuations $\sigma,$
the deviations from the Gaussian distribution are stronger at high redshifts $z$. It might naively 
be expected that as the fluctuations grow, the PDF becomes more non-Gaussian.
However, at first sight, the reverse occurs. For example, in the
top panels of Figure~\ref{fig:Fig7} (evolutionary tracks), 
the factors $S_3$ and $S_4$  {decrease} with decreasing redshift at fixed $\sigma$. 

\begin{figure}[ht] 
\centering 
\hspace{2mm}
\resizebox{0.45\textwidth}{!}{\includegraphics*{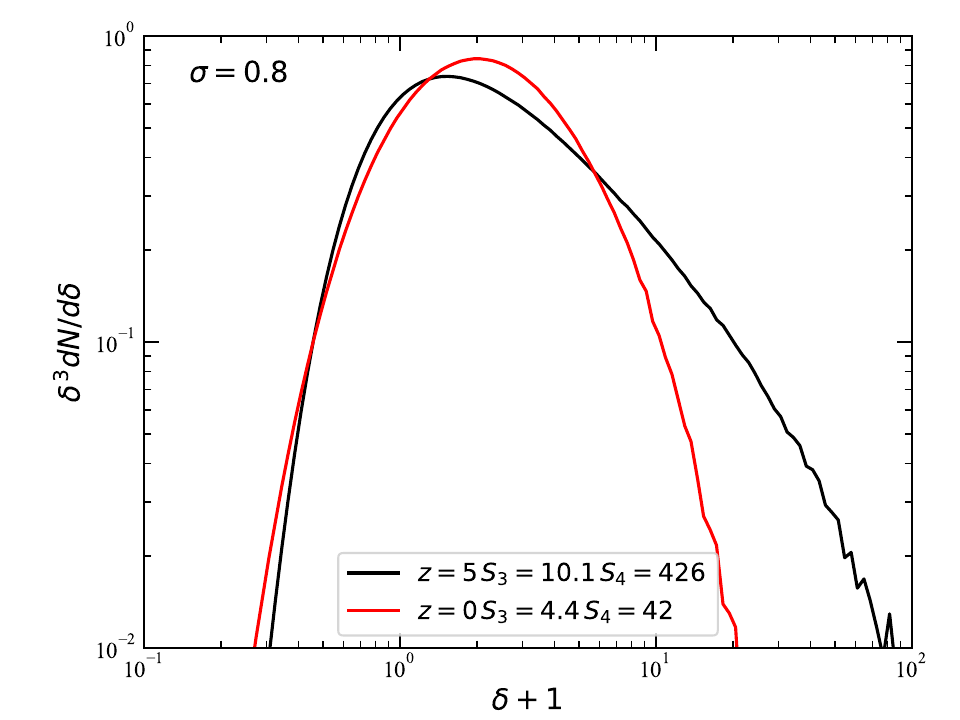}}
\caption{Comparison of the PDFs with the same rms fluctuations $\sigma=0.8,$ 
but estimated at different redshifts $z=0$ and $z=5$. 
The PDFs are scaled with $\delta^3$ to reduce the dynamical range.
The filtering scales are 
 $R_t=10~\Mpc$ at $z=0$ and $R_t=0.5~\Mpc$ at $z=5$. } %
\label{fig:Fig9} 
\end{figure}

In order highlight this effect, in Figure~\ref{fig:Fig9} we plot the PDFs
selected at two different redshifts that have the same $\sigma=0.8$.
It is clear that at $z=5,$ the PDF is much broader and more evolved
than at $z=0$.  The key issue here is that the filtering scale $R_t$
is dramatically different for $z=5$ and $z=0$.  In order to
have the same $\sigma$ at high redshift, the
filtering scale needs to be decreased.  In our case, this amounts to changing $R_t$ from
$10~\Mpc$ at $z=0$ to $0.5~\Mpc$ at $z=5$.
Or in different
terms: at redshift $z=5,$ small-scale structures of the cosmic web,
highlighted by smoothing with $R_t=0.5~\Mpc$, are more asymmetric than
supercluster-scale structures at the present epoch.

This effect becomes quite obvious when we understand why it
occurs. However, it presents a problem for non-linear models of the
PDF, 
such as the lognormal distribution where the rms of the density
perturbation $\sigma$ is the only factor that defines the PDF. These
models cannot possibly account for the evolution of the PDF with
redshift.

{ The predictions of the PT calculated with Eqs.~(\ref{eq:PT}) and presented
  in Figure \ref{fig:Fig6} suggest the dependence of $S_3$ and $S_4$
  on the effective slope of the power spectrum of the perturbations at
  radius $R$ as $\gamma=-(n_{\rm eff}+3)$.  This only affects the
  height of values $S_3$ and $S_4$, see the ladder of dotted lines in
  the top panels of Fig. \ref{fig:Fig6} for different $R_t$ and the lines in bottom
  panels of Fig. \ref{fig:Fig6} for different $z$.  No strong increase in
  $S_3$ and $S_4$ for later evolutionary epochs, $z \le 10$, and
  smaller smoothing scales, $R_t \le 4$, is predicted.  }

\begin{figure}[ht] 
\centering 
\hspace{2mm}
\resizebox{0.45\textwidth}{!}{\includegraphics*{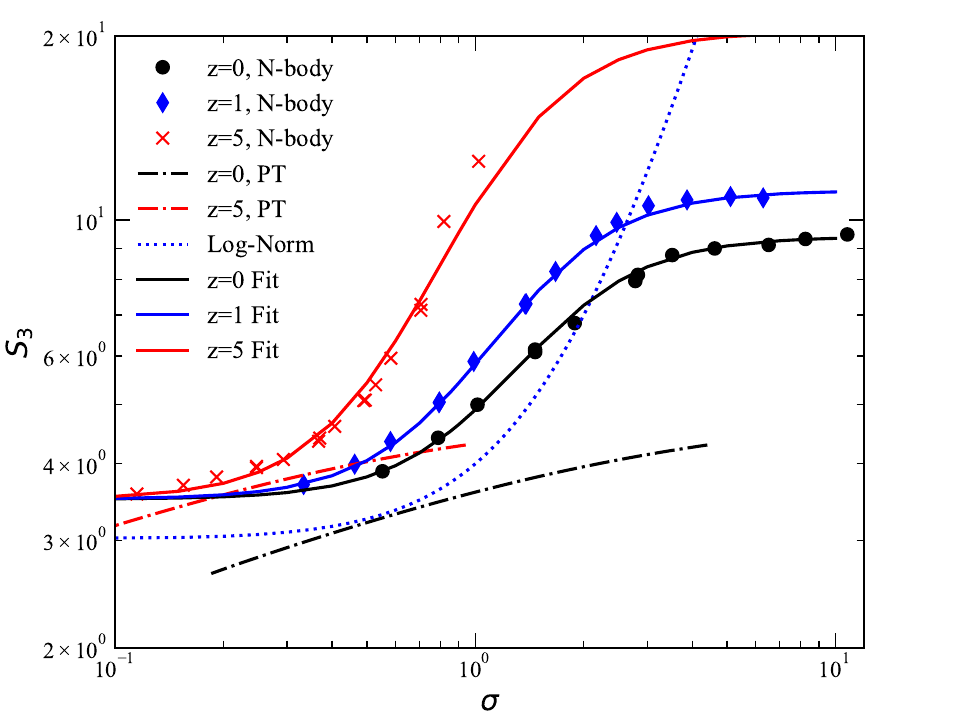}}
\caption{Comparison of $S_3(\sigma,z)$  for
  redshifts $z=0, 1, \text{and } 5$ found in $N-$body simulations with
  the predictions of the PT, eqs.~(\ref{eq:PT}), and the log-normal
  distribution, eqs.~(\ref{eq:LN}).  } %
\label{fig:Fig10} 
\end{figure}

In Figure~\ref{fig:Fig10} we compare the evolution of $S_3(\sigma)$ of
the GLAM simulations with results of the PT, the lognormal
distribution, and an analytical fit. { The new analytical
  approximation  is described
in the next subsection.}  The  fit uses as
input $\sigma$ and redshift $z$ and gives $S_3$ and $S_4$ using an approximation of four 
free parameters.  It works well and can be
useful for predictions. The log-normal and PT results are not very
accurate. The log-normal distribution does not have any dependence on
redshift because by design, it is a function of $\sigma$ only. It fails
on all scales, small and large $\sigma,$ and even at high redshifts.
The PT approximation predicts some evolution with redshift, but the
magnitude of the effect is just too small.

{\scriptsize 
\begin{table*}[ht] 
  \caption{Estimates of the  skewness  and kurtosis  parameters}
\centering
\begin{tabular}{lccccc}
\hline  \hline
  Author & Model& Box size $L_0$   & Redshift interval & $R_t$&
                                                              Moments\\
  && $\Mpc$ & & $\Mpc$& \\
  \hline  \\
\citet{Kofman:1994aa} & SCDM & 200 & $0.5 \ge a \ge 1.0$& $5 - 21$&$S,~K$\\
\citet{Romeo:2008aa}&$\Lambda$CDM &20&$15 \le z \le 0$&$\ll 1$ &$S,~K$\\
\citet{Hellwing:2010aa}&SCDM,$\Lambda$CDM, ReBEL& 180 - 1024&$ 5 \le z \le
                                                        0$& $1 - 100$& $S_3  - S_8$\\
  \citet{Mao:2014aa}     & $\Lambda$CDM &2400& $z=0$& $10 - 100$&$S_3,~S_4$ \\
\citet{Shin:2017aa}       &$\Lambda$CDM& 1024&$4 \le z \le 0$& $2 - 25$&$S_3,~S_4$ \\
\citet{Hellwing:2017aa}&$\Lambda$CDM,~nDGP&1024&$1 \le z \le 0$& $1 -
                                                                 100$& $S_3  - S_8$\\
  This work                      &$\Lambda$CDM& 256 - 4000& $30 
                                                           \le   z \le   0$& $ 0.4 - 32$&$S,~K,~S_3,S_4$
\label{Tab3}                         
\end{tabular} 
\end{table*} 
}

\subsection{Analytical approximations for the evolution of skewness and kurtosis} 

We fitted the results of the evolution of $S_3(\sigma)$ and $S_4(\sigma)$ using a four-parameter functional form.
If $S$ is either $S_3$ or $S_4$, then the approximation can be written as
\be
S = S_{\rm min} + \frac{C_1\sigma^{5/2}}{1+C_2\sigma^{5/2}},
\label{eq:Sfit}
\ee
where $S_{\rm min}$ is the minimum value of  $S$, and parameters $C_1$
and $C_2$ describe the shape of the function $S(\sigma)$. 
Specifically, the ratio $C_1/C_2$ gives the magnitude of the total increase in $S,$ 
\be
C_1/C_2 = S_{\rm max} - S_{\rm min}.
\ee
Another shape parameter is $S_1$, which is the value of $S$ at
$\sigma=1$. Using parameters $S_{\rm min}$, $S_1$, and $S_{\rm max}$
, we can find 
$C_1$ and $C_2$,
\be
C_1 =\frac{(S_1-S_{\rm min})(S_{\rm max}-S_{\rm min})}{(S_{\rm
    max}-S_1)}, \quad C_2=\frac{C_1}{(S_{\rm max}-S_{\rm min})}. 
\label{eq:Cpars}
\ee
The evolution of $S$  is described by a power-law dependence of $S_1$ and $S_{\rm max}$ on redshift $z$,
\be
S_1 = S_{10}(1+z)^\alpha, \quad S_{\rm max} = S_{\rm max,0}(1+z)^\alpha,
\ee
where $\alpha$ is a free parameter describing the evolution with
time. Here $S_{10}$ and $S_{\rm max,0}$ are shape parameters estimated
at $z=0$. 

This approximation has four free parameters: three shape parameters
$S_{\rm min}$, $S_{10}$, and $S_{\rm max,0}$ , and parameter $\alpha,$
 which describes the evolution. 
Based on these parameters, $C_1$ and $C_2$ can be determined using
Eq.(\ref{eq:Cpars}). Now we can use Eq.(\ref{eq:Sfit}) to estimate
parameters $S_3$ and $S_4$. 
For the cosmology used for the {\sc GLAM} simulations and for a reduced skewness $S_3$ , the parameters are
$S_{3,\rm min}=3.5$, $S_{3,10}=4.9$, and $S_{3\rm
  max,0}=9.4$. Parameter $\alpha$ depends on redshift. We find that
for $z\lesssim 1.5,$  
the parameter is $\alpha=0.25$. For higher redshifts, the evolution is
slightly faster: $\alpha=0.43$. These approximations give a 5\%
accuracy for $S_3$. 
This fit was tested and can only be used for $z\lesssim 20$ and for
$\sigma >0.05$.

\subsection{Comparison with earlier results}

{ Only a few studies exist of the moments of the PDF on the basis of
numerical simulations of the evolution of the density field.
We summarise the results from different simulations in
Table~\ref{Tab3}. Authors have used different cosmological parameters in
simulations and various redshift intervals and smoothing lengths. A
direct comparison of the skewness and kurtosis parameters is not easy.

 Only a few authors provided results of the
  evolution of the $S_3$ and $S_4$ parameters in cosmological $N$-body
  simulations, and none of them provided data for the dependence on the
  amplitude of the density fluctuations $\sigma$.  \citet{Shin:2017aa}
  showed results for the evolution of the $S_3$ parameter for a top-hat
  filter with radius $R_{\rm TH}=10~\Mpc$. Our results are very
  similar to theirs for the same effective volume, although there are
  differences on a level of a few percent. \citet{Mao:2014aa} showed
  results for $z=0$ and $R_{\rm TH}=10~\Mpc$, which also agree with
  our results within $\sim 10\%$  errors.

  There is a reason for the lack of interest in the evolution of the
  PDF with time:  it is expected that the PDF depends on time only through
 the amplitude of perturbations $\sigma$. In other words, it is expected
 that $P(\rho) = P(\rho; \sigma(z,R))$.  Perturbation theory
  \citep[e.g.][]{Juszkiewicz:1993aa, Bernardeau:1994aa} predicts some 
  explicit dependence on $z,$ as can been seen in Eqs.~(\ref{eq:PT}).
   However, the dependence on redshift is very weak. At the same time,
   non-linear approximations based on top-hat collapse or spherical
 infall models \citep[e.g.][]{Betancort-Rijo:2002ve,Lam:2008vy} only depend
  on $\sigma$ and not explicitly on the redshift. 

 As our results clearly show, an explicit evolution with the redshift is
 clearly present and quite strong.  This suggests
  that some presumptions, on which the PT is based, need revision. 
 }

\subsection{Cosmological interpretation}

\subsubsection{Contrasting the evolution of the cosmic web on small and
  large scales}

One of the findings of our study is the contrast between the evolution
of the cosmic web on small and large scales, as defined by the smoothing
length $R_t$.   The cosmic web populations defined by a large smoothing
length $R_t \ge 10~\Mpc$, at all cosmic epochs have $S_3 \le 4$ and
$S_4 \le 35$. The cosmic web populations defined by a small
smoothing length $R_t \le 2~\Mpc$, at late evolutionary epochs
$z \le 5$ have maxima of moments $S_3 \ge 10$ and $S_4 \ge 150$.

To understand the reason for this difference, we recall that
the cosmological  moments $S_3$ and $S_4$ are actually amplitude parameters of
the mathematical skewness $S = S_3 \times \sigma$ and kurtosis
$K= S_4 \times \sigma^2$, as defined by Eqs.~(\ref{s3c}) and
(\ref{s4c}).  The data presented in Figs.~\ref{fig:Fig6} and
\ref{fig:Fig7} can be expressed as functions of redshift of the
mathematical skewness $S(z)$ and kurtosis $K(z)$, as done in
Figure~\ref{fig:Fig3}.  The evolution of large-scale populations
of the cosmic web proceeds with an almost constant rate that is measured by the mean
logarithmic gradients, $\langle\gamma_S\rangle \approx -1$ and
$\langle\gamma_K\rangle \approx -2$, shown in Figure~\ref{fig:Fig4}.
The speed of the evolution of the small-scale elements, characterised by
a small smoothing length $R_t$, is much faster: for $R_t=1~\Mpc$
$\langle\gamma_S\rangle \approx -1.5$ and
$\langle\gamma_K\rangle \approx -3$, see Figure~\ref{fig:Fig4}.

The increase in the (negative) gradient in the interval $z \le 10$ is
due to the non-linear growth of that density perturbations, which is important for
small-scale perturbations.  The relative decrease in the gradient for
$z \le 1$ is due to the effect of the $\Lambda$ term.

\subsubsection{Similarity of the evolution of the skewness and kurtosis}

We note one important property of the PDFs of the cosmic web: the shapes
of the $S_3(\sigma)$ and $S_4(\sigma)$ curves are qualitatively very
similar, as shown in Figure~\ref{fig:Fig6} and especially in
Figure~\ref{fig:Fig7}.  This property is due to the character of the density
field of the cosmic web. It is highly asymmetric, all details of the
structure are in over-density regions, and the under-density region is
almost structure-less. We conclude that the two PDF moments, the skewness $S(\sigma)$
and the kurtosis $K(\sigma)$ (and their amplitudes $S_3(\sigma)$ and
$S_4(\sigma)$), essentially measure the asymmetry of the density field. 
However,  large quantitative differences exist:   at
the maxima, $S_4(\sigma)$ are larger than $S_3(\sigma)$ by a factor of
10 to  50, see Fig.~\ref{fig:Fig8}.

\subsubsection{Independent evidence for the asymmetry of the density
  field}

The asymmetry of the evolution of the density perturbations is reflected
not only in the PDF of the density field as measured by the skewness
parameter $S$.  It is also seen in the distribution of particle
densities, as shown by \citet{Pandey:2013aa}.  Asymmetry is also observed in the distribution of the number of superclusters as a function of
the reduced density, $\nu=\delta/\sigma$, see Figure~1 by
\citet{Einasto:2019fk} and Figure~2 by \citet{Einasto:2021aa}.

According to our assumptions, the evolution of the universe started from a Gaussian 
random field that was symmetrical around the
mean density, that is, {  positive and negative deviations from the mean
density are equally probable}. The question thus is at which time the density field became asymmetric. 

To explore the problem, we investiage the evolution of
the structure shown in Figure~\ref{fig:Fig1}.
For example,  supercluster-type elements are visible almost in the
same form already at the earliest epoch, $z = 30$, and they change little
at the late stage of the evolution. Cluster-type elements are also seen in 
the early universe, but they change much during the evolution.  These differences
illustrate the numerical data of the evolution on various scales, shown in
Figure \ref{fig:Fig6}.  For our study, the
presence of both small- and large-scale elements of the cosmic web
already at early stages of the evolution is important. 

Another
manifestation of the early evolution of the cosmic web is the almost
constant number of superclusters during the evolution, see Figure~2 in
\citet{Einasto:2019fk} and Figure~6 in \citet{Einasto:2021aa}. This
suggests that supercluster embryos were created in the very early
universe, much earlier than is seen in the density field of the cosmic
web at redshift $z=30$.  The difference between positive and negative
density perturbations lies in the fact that positive perturbations
form distinct structures, embryos of galaxies, clusters, and
superclusters, already at the very early stages of the evolution,
whereas negative perturbations of similar strength form voids and act as
structure-less repellers.  The asymmetry parameter, the mathematical skewness $S$, measures
this difference in positive and negative perturbations.

\subsubsection{PDF moments in the  early universe}

The behaviour of the skewness and kurtosis at very early
epochs is of interest. Figs.~\ref{fig:Fig6} and \ref{fig:Fig7} show that all
$S_3(\sigma)$ and $S_4(\sigma)$ curves approach with increasing $z$
limiting values, depending on the scale of systems, as 
determined by smoothing length.   In this way, our analysis
confirmed earlier results by \citet{Bernardeau:1995aa},
\citet{Hellwing:2010aa,Hellwing:2017aa}, and \citet{Mao:2014aa}.  

We cannot answer the question how PDF moments $S(\sigma)$,
$K(\sigma)$, $S_3(\sigma)$ and $S_4(\sigma)$ behave at very small
$\sigma$\,at the moment.  The cosmic density field can have small rms of density
fluctuations $\sigma$ in a young universe at high redshift $z$, or
using a very large smoothing length $R_t$.  Our data suggest that in a
young universe, the PDF moments converge with increasing $z$ to limits
$S(\sigma)= S_3\,\sigma$, and $K(\sigma) = S_4\,\sigma^2$ with
$S_3 \approx 3$ and $S_4 \approx 15$. The limited range of the smoothing
lengths $R_t$ used in our simulations gives no hint to $S_3(\sigma)$
and $S_4(\sigma)$ for very large smoothing. Future studies are needed
to solve this question.

\subsubsection{Early  evolution of the universe}

The early evolution of the density field was calculated in simulations
using the Zeldovich approximation, thereafter, actual numerical
simulations follow.  Available data suggest that embryos of galaxies
and superclusters were created by high peaks of the initial field.
The initial velocity field around the peaks is almost laminar.  The highest
density peaks of the density field started to attract surrounding
matter more strongly than around peaks of lower density.  In this way,
centres of future galaxies, clusters, and superclusters formed.  The
almost identical pattern of the cosmic web on the supercluster scale at epochs with
$z \ge 3$ suggests that the same pattern existed at earlier epochs,
even soon after the creation of density fluctuations.  The
development of the density field in the early phase is well described
by the \citet{Zeldovich:1970} approximation and its extension, the
adhesion model by \citet{Kofman:1988aa}.  As shown by
\citet{Kofman:1990aa, Kofman:1992aa}, the adhesion approximation
for the present epoch yields structures that are very similar to the structures
calculated with N-body numerical simulations of the evolution of the
cosmic web with the same initial fluctuations. {\em \textup{Thus the combination
of theoretical models and numerical simulations suggests that the
asymmetry of the PDF started to form soon after the creation of
fluctuations in the early period of the evolution of the
universe.} }

\section{Conclusions}

We studied the evolution of the DM density field with the goal of determining
evolutionary changes in one-point PDF and its moments.  We used a large
set of input parameters of $N$-body simulations, the box size,
$L_0$, the smoothing length, $R_t$, and the simulation epoch, $z$, to follow
the evolution of $\Lambda$CDM models.  We performed numerical
simulations of $\Lambda$CDM models for two sets of simulations, one
with $N_{\mathrm{grid}} = 512$, and the other with
$N_{\mathrm{grid}} = 2000,~5000$.  In these sets we used different
cosmological parameters, simulation algorithms, and simulation box
sizes.  We calculated density fields for several series of smoothing
lengths using various smoothing rules.

For all simulation sets, we calculated one-point PDFs and their moments, the
standard deviation $\sigma$, the skewness $S,$ and the kurtosis $K$.  { The mathematical
  skewness $S$ characterises the degree of asymmetry of the
  distribution, while the kurtosis $K$ measures the presence of heavy
  tails and peaks in the distribution. Simple relations
exist  between mathematical and cosmological parameters: the skewness 
  $S = S_3 \times\sigma$, and the kurtosis $K = S_4 \times\sigma^2$, where
  $S_3$ and $S_4$ are the cosmological skewness and kurtosis, that is,
the   cosmological skewness and kurtosis, $S_3$ and $S_4$, play the role
  of amplitude parameters of the mathematical $S$ and $K$.  }

Our study extends
previous studies by analysing both mathematical and cosmological skewness
and kurtosis, using a wide range of evolutionary epochs from $z=30$
on, and a wide range of smoothing lengths from $R_t=0.4$ to
$R_t=32~\Mpc$.  We defined populations of the cosmic web by the
smoothing length $R_t$, which is used to calculate PDF moments.

The basic conclusions of our study are listed below.
 
\begin{enumerate}

\item{} The moments $S$ and $K$, calculated for density
  fields at different cosmic epochs and smoothed with various scales,
  characterise the evolution of different structures of the web.
  The moments calculated with small-scale smoothing
  ($R_t\approx 1-4~\Mpc$) characterise the evolution of the web on
  a cluster-type scale.  The moments found with large smoothing
  ($R_t\gtrsim 5-15~\Mpc$) describe the evolution of the web on
  a supercluster scale.  

\item{} During the evolution, the cosmological  skewness
  $S_3= S/\sigma$ and cosmological  kurtosis $S_4=K/\sigma^2$ present a
  complex behaviour: at a fixed redshift, the curves of $S_3(\sigma)$ and
  $S_4(\sigma)$ steeply increase with $\sigma$ at $\sigma\lesssim 1$
  and then flatten out and become constant at $\sigma\gtrsim 2$.  If
  we fix the smoothing scale $R_t$, then after reaching the maximum at
  $\sigma\approx 2$, the curves at large $\sigma$ start to gradually
  decline. We provided accurate fits for the evolution of
  $S_{3,4}(\sigma,z)$.  Skewness and kurtosis approach
   constant levels at early epochs: depending on the smoothing
  length, $S_3(\sigma) \approx 3$ and
  $S_4(\sigma) \approx 15$, respectively. 
 
\item{} Direct and indirect data suggest that seeds of elements of the
  cosmic web were created at early epoch at inflation and started to
  grow thereafter.  This explains the continuous growth of the asymmetry of
  the density distribution, expressed by the  skewness $S$
  and kurtosis $K$ functions.

\end{enumerate}

We find that 
the evolution of $S_3$ and $S_4$ cannot be described by current theoretical 
approximations. The often-used lognormal distribution function for the PDF fails
to explain even qualitatively the shape and evolution of $S_3$ and $S_4$. 
We still have no definite answer to
the question how the PDF moments behave at very small $\sigma$. The
cosmic density field has a small rms of the density fluctuations in a young
universe and in the universe at the present age, applying a very large
smoothing length of the density field.  The limited range of the smoothing
lengths used in our simulations gives no hint to $S_3$ and $S_4$ for
very large smoothing.

\begin{acknowledgements}
  
We thank Ivan Suhhonenko for performing GADGET simulations, used in
this study,    Enn Saar for discussion, and
anonymous referee for very stimulating suggestions, which helped in
improve the paper.
  This work was supported by institutional research funding IUT40-2 of
  the Estonian Ministry of Education and Research, by the Estonian
  Research Council grant PRG803, and by Mobilitas Plus grant MOBTT5. We
  acknowledge the support by the Centre of Excellence``Dark side of
  the Universe'' (TK133) financed by the European Union through the
  European Regional Development Fund.

\end{acknowledgements}


\begin{appendix}

\section{Density estimates and power spectra}
  
\subsection{Kernel density estimates}
\label{app:kern}

One of the ways to calculate the density field is through a kernel sum \citep{davison1997},
\begin{equation}
    \rho(\mathbf{r}) = \sum_{i=1}^N K(\mathbf{r} - \mathbf{r}_i),
    \label{eq:dens}
\end{equation}
where the sum is over all $N$ data points, $\mathbf{r}_i$ are the
coordinates of data points, and $K$ is the kernel.  

Kernels $K$ are required to be distributions that are positive everywhere and
integrate to unity; in our case, 
\begin{equation}
    \int K(\mathbf{y})d^3y=1.
    \label{eq:kern}
\end{equation}
Good kernels for smoothing densities to a grid are the box splines
$B_k$. They are local, and they are 
interpolating on a grid,
\begin{equation}
    \sum_i B_k \left(x-i \right) = 1,
    \label{eq:sum}
\end{equation}
for any $x$ and a small number of indices that give non-zero values
for $B_k(x)$. To create our density fields, we used the popular $B_3$
spline function,
\begin{equation}
    B_3(x) = \frac{|x-2|^3 - 4|x-1|^3 + 6|x|^3 - 4|x+1|^3 + |x+2|^3}{12}.
\end{equation}
This function differs from zero only in the interval $x\in(-2,2)$,
meaning that the sum in (\ref{eq:sum}) only includes values of
$B_3(x)$ at four consecutive arguments $x\in(-2,2)$ that differ by 1.
Figure~\ref{fig:tuum} shows the shape of the function in comparison to
a Gaussian.  We note that in this formulation, the
smoothing scale and input particle coordinates are given implicitly in units of
grid cell length, and the smoothing scale is equal to one.

If we choose $B_3$ to be our kernel function, $K_B^{(1)}(x) = B_3(x)$,
then the three-dimensional kernel $K_B^{(3)}$ is given by a direct
product of three one-dimensional kernels,
\begin{equation}
    K_B^{(3)}(\mathbf{r}) \equiv K_B^{(1)}(x)\, K_B^{(1)}(y)\, K_B^{(1)}(z),
    \label{eq:3dkern}
\end{equation}
where $\mathbf{r} \equiv \{x,y,z\}$. Although this is a direct
product, it is isotropic to a very high degree 
\citep{saar2009}.

\begin{figure}[ht] 
\centering 
\hspace{2mm}
\resizebox{0.45\textwidth}{!}{\includegraphics*{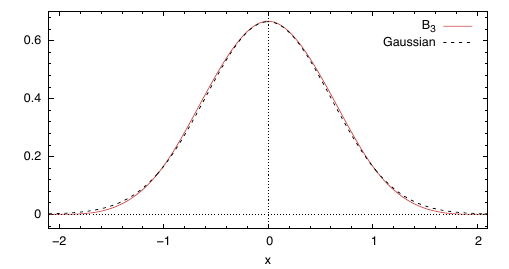}}
    \caption{\footnotesize Shape of the box spline kernel
      $B_3$. The solid curve shows the $B_3(x)$ kernel, and the dashed curve shows a
      Gaussian with $\sigma=0.6$.} 
    \label{fig:tuum}
\end{figure}

To increase the smoothing scale, we can introduce a scale parameter to
the kernel function $K$ (together with the appropriate normalisation).
However, in practice, this also increases the number of particles in the
kernel volume, and the computation can quickly become uneconomical.
One of the benefits of using the $B_3$ smoothing kernel is that we
can employ the à trous wavelet algorithm, and having an existing
density field as basis, calculate a field with twice the smoothing
length by convolution with a simple discrete filter
\citep{starck2006},
\begin{equation}
    \begin{array}{rcl}
    C_{j+1}(i_x,i_y,i_z) = \displaystyle{\sum_{l,m,n}} H(l,m,n)\,
    C_j(i_x+2^{j}l,i_y+2^{j}m,i_z+2^{j}n).\hspace{-10pt}\\[30pt]
    \end{array}
    \label{eq:atrous}
\end{equation}
Here $C_j$ and $C_{j+1}$ denote correspondingly the initial and
smoothed density fields.  
Thus only the first density field needs to be calculated using particle coordinates.
The convolution mask $H$ is constructed as the following direct product:
\begin{equation}
    H = h\otimes h\otimes h,
\end{equation}
where the coefficients $h$ are derived from the à trous discrete
wavelet transform, and its values corresponding to the $B_3$ function are 
\begin{equation}
    h = \left\{\frac{1}{16};\frac{1}{4};\frac{3}{8};\frac{1}{4};\frac{1}{16}\right\}.
\end{equation}

\subsection{Cubic-cell   density estimate}

Smoothing lengths of original density fields from numerical simulation
output have a resolution that is equal to the size of the simulation cell,
$L_0/N_{\mathrm{grid}}$. We call this the smoothing rank zero. We used a
smoothing recipe that increased the smoothing length by a factor of
2.   We used this recipe successively four times, and obtained four smoothing
ranks 1 to 4.  For clarity, we show in the core text a smoothing
length in units of $\Mpc$.

In case of the cubic-cell smoothing, we divided the original
computational box of size $L_0$ with resolution
$N_{\mathrm{grid}} = 512$ to boxes of smaller resolution,
$N_{\mathrm{grid}} = 256,~128,~64,~32$, by counting densities in
respective cells of the field in the previous smoothing length.  This
yielded smoothing lengths $R_t = L_0/N_{\mathrm{grid}}~\Mpc$.
Smoothing lengths for rank 1 are $R_1=L_0/256~\Mpc$,
$R_2=L_0/128~\Mpc$ for rank 2, and so on.  The main difference
between smoothing rules is that $B_3$ spline and top-hat rules
preserve the grid size, but the cubic-cell rule yields density fields with
decreasing grid sizes $N_{\mathrm{grid}}$.

\begin{figure*}[ht] 
\centering 
\hspace{2mm}
\resizebox{0.75\textwidth}{!}{\includegraphics*{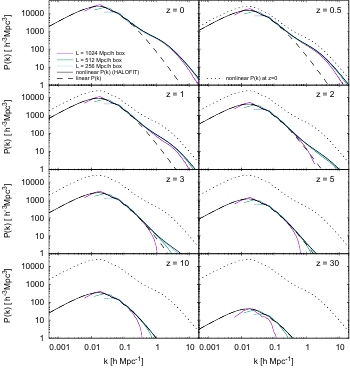}}
\caption{ Comparison of simulation power spectra with the
  corresponding linear spectra and nonlinear HALOFIT results at
  various epochs $z$.  }%
\label{pk_app} 
\end{figure*} 

\subsection{Top-hat density and error estimates}
\label{app.error}

Top-hat smoothed densities for L256, L512 and L1024 simulations were
calculated at the nodes of a $512^3$ regular cubic grid using
smoothing radii for ranks 1 to 4, as explained in the previous subsection.
 
The errors for the moments of the density distribution were determined
through jackknife resampling: a full simulation box was divided into
smaller cubes, each time omitting one of the small cubes while
calculating the statistical moments. The variability of the calculated
moments directly gives the desired error estimates
\citep{Efron:1982aa}. We chose to split the simulation box into $2^3$,
$4^3$ , and $8^3$ equal-sized subcubes. The error estimates for all of
these three choices did not vary significantly. In figures with error,
we only use the errors for the second choice, that is, $4^3 = 64$ subvolume
case.  We note that errors of all quantities (variance, $K$,
$S$, $S_3$, $S_4$) were found separately.

The PDFs of density fields and their moments can have systematic
errors.  We discuss these errors in the next appendix.

\subsection{Evolution of the DM power spectra}

Here we perform a simple consistency check for our {\sc GADGET} simulations. In
particular, we calculate the density power spectra for all the output
redshifts and box sizes, resulting in a total of $8\times 3=24$
spectra. The spectra were calculated using FFTs on a $1024^3$ regular
cubic grid. The simulation particles were assigned to a grid through the
triangular-shaped cloud (TSC) mass-assignment scheme. The shot-noise
removal, grid smoothing, and aliasing correction tailored for the
TSC scheme were applied following the description given
by~\citet{Jing:2005wv}.

The results of these calculations are shown in Figure~\ref{pk_app}. Here
the panels correspond to simulation snapshot redshifts displayed in
the upper right corners, that is, they grow from left to right
and from top to bottom. In comparison, we show the non-linear spectra according to the analytic HALOFIT
approximation~\citep{Smith:2003ui,Takahashi:2012tg} as
implemented in the cosmological Boltzmann code package
CAMB\footnote{https://github.com/cmbant/CAMB} by \citet{Lewis:2000wx}. The
corresponding linear spectra (also obtained with CAMB) are plotted as well. As a reference, the dotted curves in all of the
panels with $z>0$ show the $z=0$ non-linear spectrum.

The agreement between our simulation spectra and the
well-tested analytic HALOFIT results is mostly very good. As expected,
the larger boxes perform better at larger scales, while their lack of
resolution at smaller scales cannot properly account for the
small-scale non-linear evolution. In conjunction, except for the
highest redshifts, the chosen three box sizes are reasonably good for
capturing the dynamics of structure formation over more than three
orders of magnitude in scale.

\section{Comparison of the moments obtained with different smoothing
  recipes} 

\begin{figure*}[ht] 
\centering 
\hspace{2mm}
\resizebox{0.30\textwidth}{!}{\includegraphics*{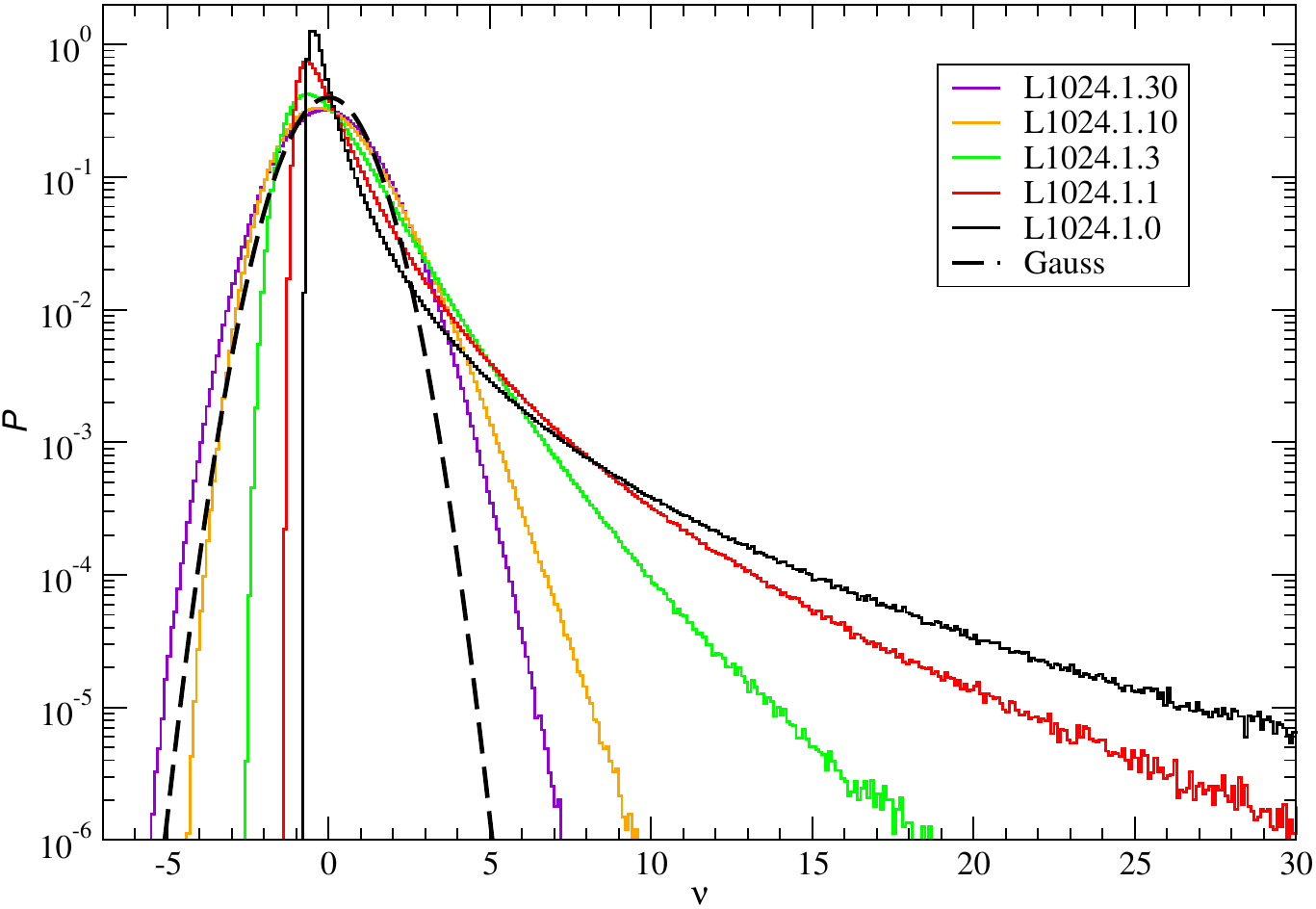}}
\hspace{2mm}
\resizebox{0.30\textwidth}{!}{\includegraphics*{L1024_S1_PDFlin.eps}}
\hspace{2mm}
\resizebox{0.30\textwidth}{!}{\includegraphics*{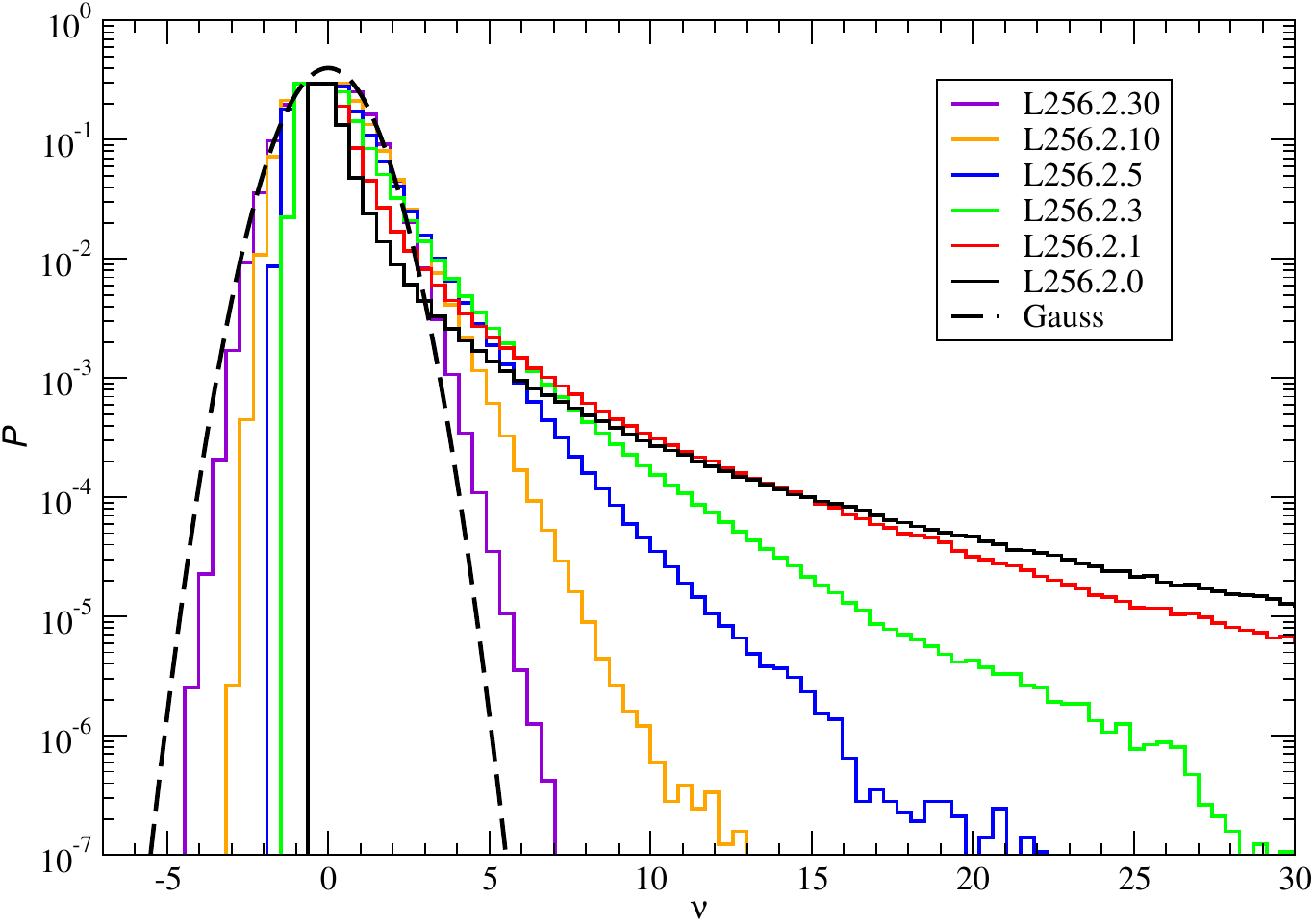}}
\caption{ PDFs as functions of reduced densities $\nu = \delta/\sigma$,
  smoothed with different methods.  {\em Left and central panels} show
simulation  L1024.1 smoothed with a $B_3$ spline and the cubic-cell  
  method, respectively, using a smoothing kernel length $R=4~\Mpc$. {\em Right panel:}
   Simulation L256.2 smoothed with the  top-hat method with
  a kernel $R=2~\Mpc$.  The smoothing rank is the first index in the
  simulation name, and the redshift is the second index. Colours indicate the
  evolutionary epoch $z=30,~10,~5,~3,~1,\text{ and}~0$.  The dashed bold curves show
  the Gaussian distribution.}
\label{fig:C1} 
\end{figure*} 

\begin{figure*}[ht] 
\centering 
\hspace{2mm}
\resizebox{0.30\textwidth}{!}{\includegraphics*{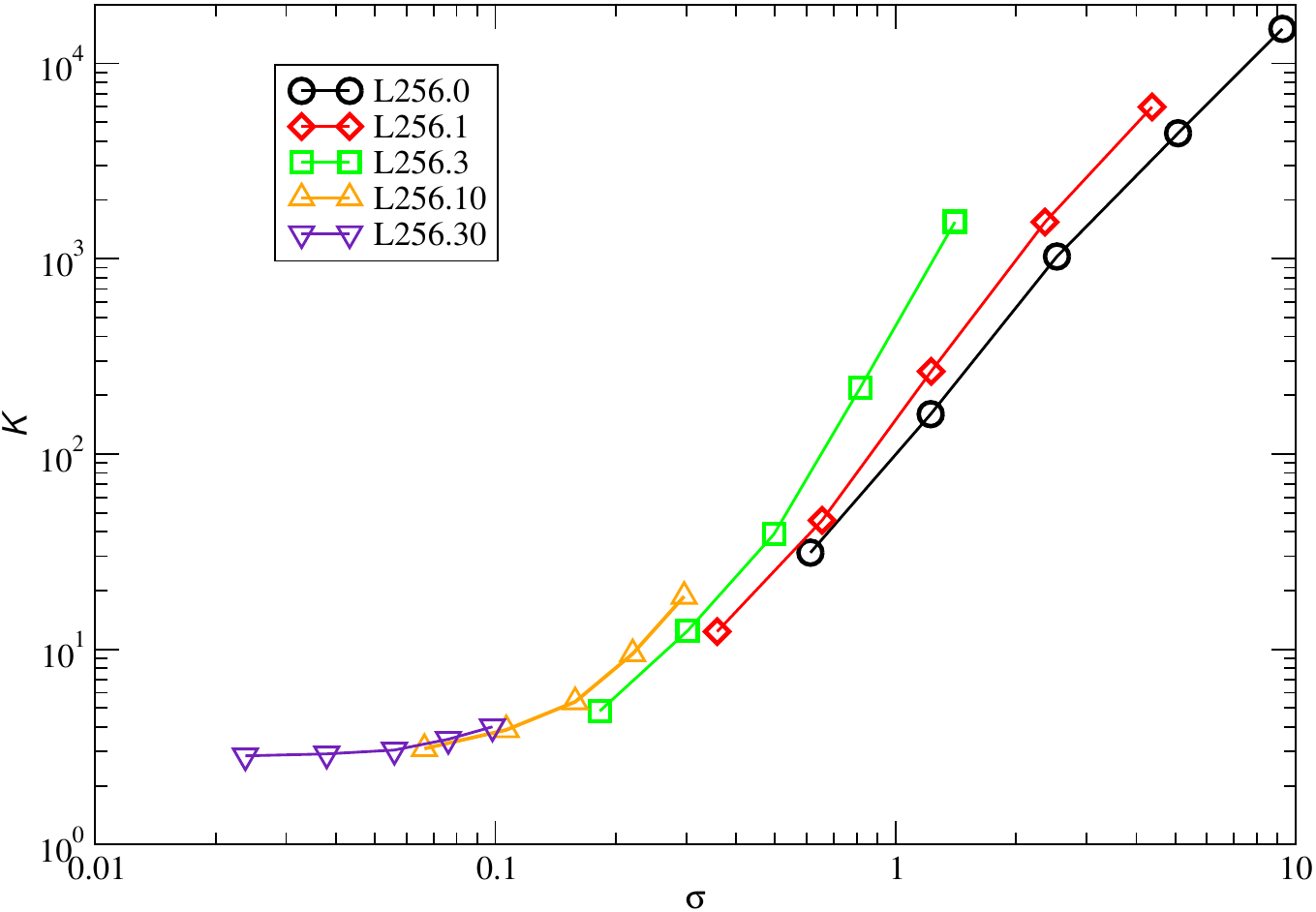}}
\hspace{2mm}
\resizebox{0.30\textwidth}{!}{\includegraphics*{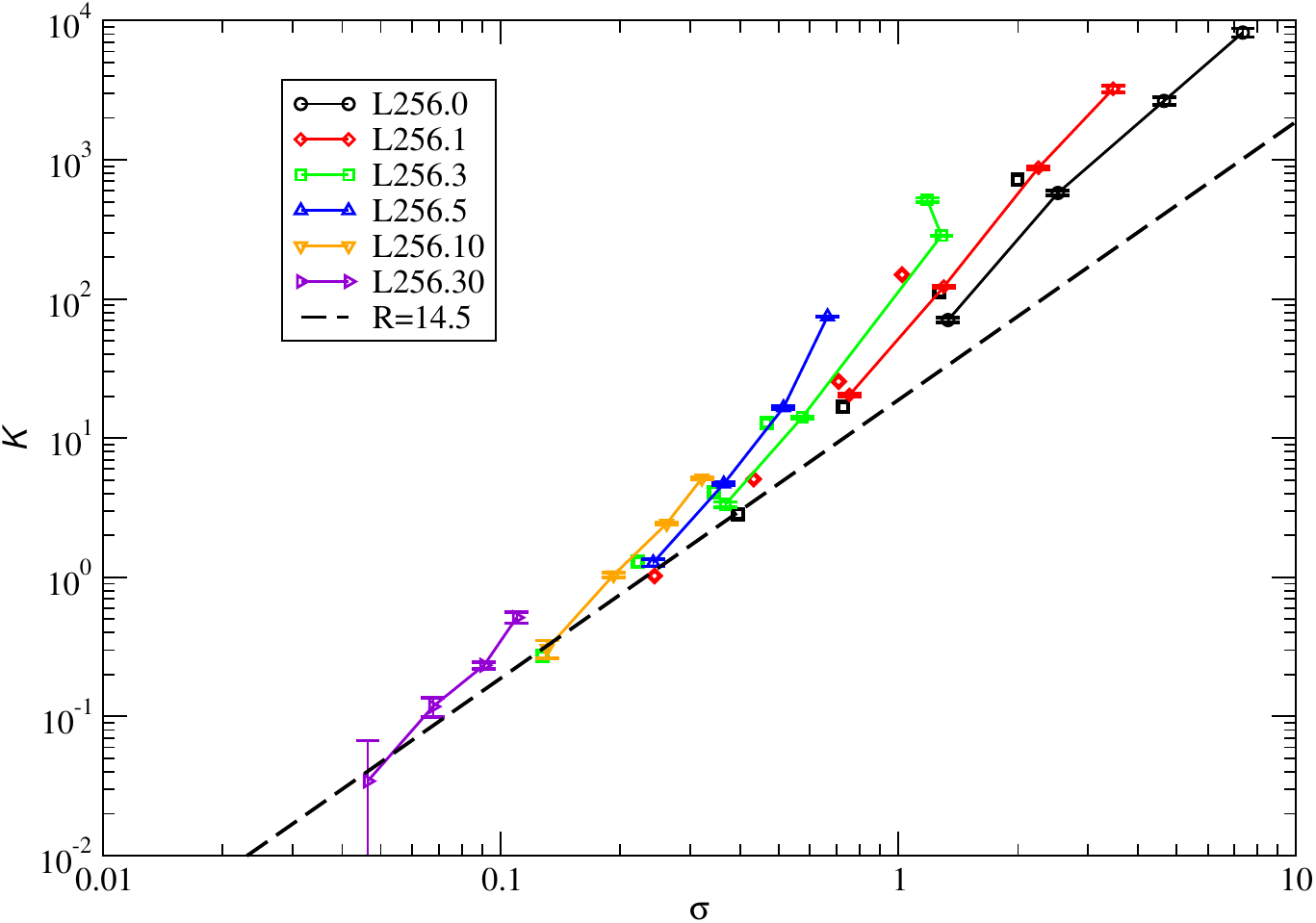}}
\hspace{2mm}
\resizebox{0.30\textwidth}{!}{\includegraphics*{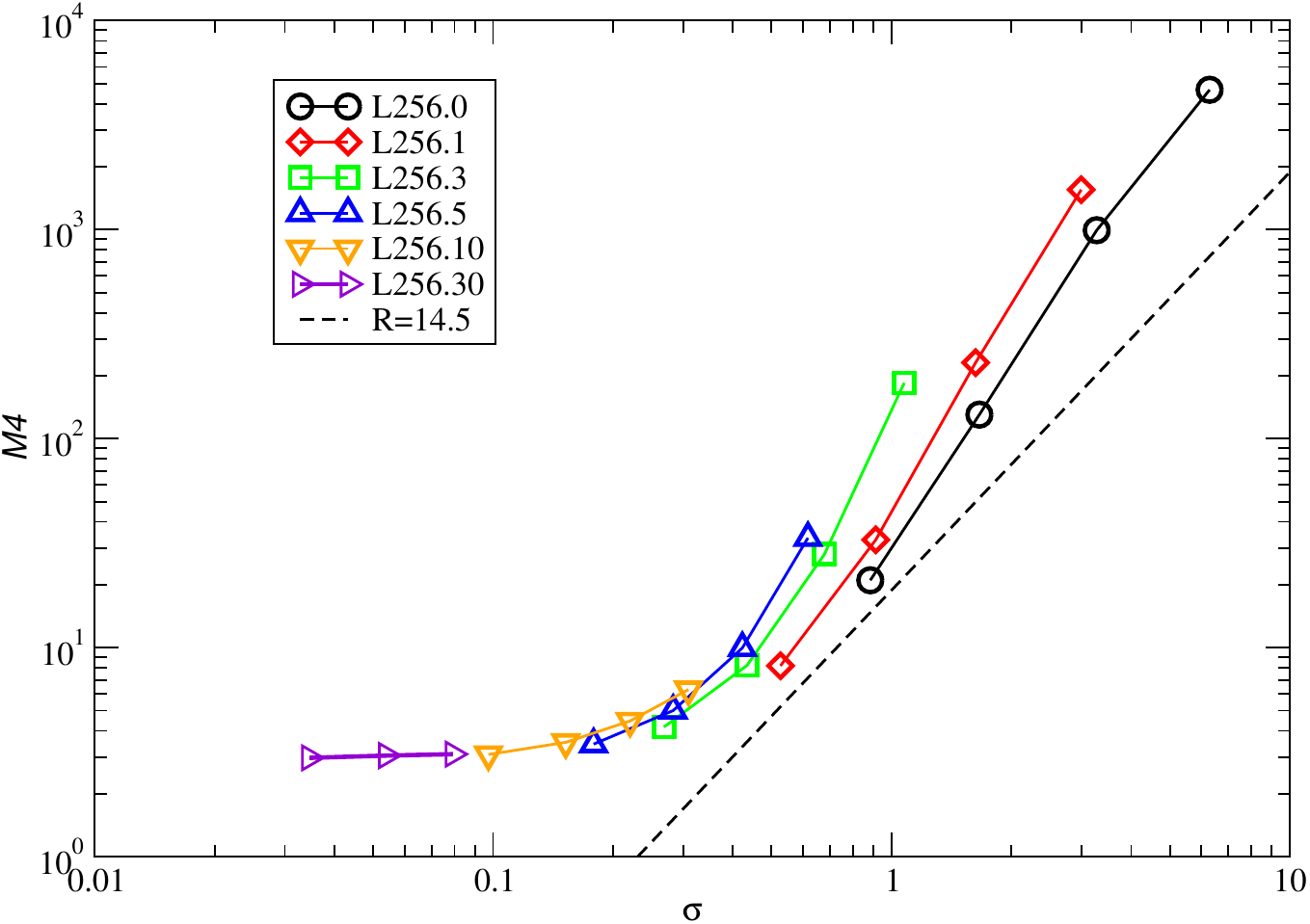}}
\caption{Mathematical kurtosis $K(\sigma)$ of simulation L256
  using various smoothing algorithms.  {\em Left panel:} Results
  for smoothing with a $B_3$ spline. {\em Central panel:} Results
  with a cubic-cell smoothing. {\em Right panel:} Results for the kurtosis
  $M_4=K+3$ using a top-hat smoothing. The name index shows the redshift. The dashed
  bold curves show  $K=16\sigma^2$.  }
\label{fig:C2} 
\end{figure*}

We smoothed our GADGET simulations with three different
kernels, the $B_3$ kernel, a cubic-cell   kernel, and a  top-hat
kernel.  All kernels yield density fields that are partly distorted
due to the insufficient resolution of our simulations for the low-density
regions. Our simulations contain one DM particle per computation
cell. During the evolution, the density in voids decreases, and there
are fewer than one particle per cell. At the present epoch, the mean
density in the central regions of the voids is about one particle per ten
cells. In the calculation of the density field, most cells in under-dense
regions contain no particles and have zero density. This distorts
the smoothed density values in under-dense regions. For this reason, we did
not use the original density fields of zero smoothing rank. Some
problems also exist when higher smoothing ranks are used, however.  In this section we
discuss some results for different smoothing kernels.

In Figure~\ref{fig:C1} we compare the PDFs obtained with different smoothing
recipes. In the left panel we show the PDFs calculated with the $B_3$
spline for simulation L1024.1, smoothed with a kernel of length $R_B=4~\Mpc$
for simulation epochs $z=0,~1,~3,~10,~\text{and }30$, given as the second index
of the sample name.  The PDF for simulation epochs
$z = 10 \text{ and } 30$ are too wide compared with the Gaussian distribution.

The central panel of Figure~\ref{fig:C1} shows the PDFs of simulation
L1024.1, smoothed with the cubic-cell   kernel of rank 1, which
corresponds to a smoothing length $R_B=4~\Mpc$.  The right panel shows
PDFs of simulation L256.2, smoothed with a  top-hat recipe
using rank 2, which corresponds to $R=2~\Mpc$.  The figure shows that
both top-hat smoothing recipes form PDFs with an expected behaviour.

Figure~\ref{fig:C2} shows the mathematical kurtosis $K$ of simulation
L256 using various smoothing recipes. The left panel shows 
the excess kurtosis of all three simulations smoothed with a $B_3$
spline. 
The coloured curves show various simulation epochs; for each epoch the  symbols
starting from top correspond to smoothing ranks 0 to 4. The figure
shows that for epochs $z = 10$ and 30, the kurtosis does not approach
the expected value for a near Gaussian distribution $K \rightarrow 0$,
but a higher value, $K \rightarrow 2.8$. In other words, for these
simulation epochs, the PDF curve is leptokurtic, that is, it  has heavy tails
on either side, as shown in the left panel of Figure ~\ref{fig:C1}.
For comparison, we present the PDF of the same simulation calculated
with the two 
top-hat recipes, shown in the central and right panels of Figure
~\ref{fig:C1}.

The central panel of Figure ~\ref{fig:C2} shows the mathematical kurtosis
$K$ for the same simulation L256, smoothed with the cubic-cell  
method. Throughout the whole $\sigma$ interval, the kurtosis $K$
is proportional to $\sigma^2$, as expected. The right panel of
Figure ~\ref{fig:C2} shows the kurtosis $M_4=K + 3$ for the  top-hat
smoothing for simulation L256. It is expected that at low
$\sigma \rightarrow 0,$ this function approaches $K(\sigma) + 3 \rightarrow 3$. The
figure shows that actually $K(\sigma) + 3 \rightarrow 2.9$. This means
that at low $\sigma,$ some vales of $K$ are negative. These deviations
are larger than the deviations expected from random errors shown in
Figure ~\ref{fig:Fig6}. In other words, here we have small, but systematic
errors.

For simulations at epoch $z=30$ and smoothing with the cubic-cell  kernel of
rank 4, we obtained slightly negative values for $S_4$ .  This is
expected because densities at this redshift are close to the mean 
density, $S_4$ is found by subtracting two approximately equal 
numbers, and the expected values of the kurtosis are lower than
the estimated errors.  In these cases, we accepted for the cosmological
kurtosis at $z=30$ the value on the basis of lower redshift, $S_4=16$,
and calculated the mathematical kurtosis $K$ using Eq.~(\ref{s4c}).

\end{appendix}

\end{document}